\documentclass{aa}

\usepackage{graphicx}

\begin{document}

\title{Optical variability of the BL Lacertae object GC 0109+224}

\subtitle{Multiband behaviour and time scales from a 7-years
monitoring campaign}

  \author{S. Ciprini\inst{1,2}, G. Tosti\inst{1,2}, C. M. Raiteri\inst{3}, M.
  Villata\inst{3}, M. A. Ibrahimov\inst{4}, G. Nucciarelli\inst{2} and L. Lanteri\inst{3}}
    \offprints{S. Ciprini \\ \email{stefano.ciprini@pg.infn.it}}
    \institute{Physics Department, University of Perugia, via A. Pascoli, 06123
Perugia, Italy
    \and
    Astronomical Observatory, University of Perugia, via B. Bonfigli, 06126 Perugia, Italy
    \and
    INAF, Osservatorio Astronomico di Torino, via Osservatorio 20, 10025
Pino Torinese, Torino, Italy
    \and
    Ulugh Beg Astronomical Institute, Academy of Sciences of Uzbekistan, Astronomicheskaya 33, Tashkent 700052, Uzbekistan
    }

 \date{Received ....; accepted ......}

 \authorrunning{Ciprini et al.}
 \titlerunning{Optical variability of GC 0109+224}

\abstract{We present the most continuous data base of optical
$BVR_{c}I_{c}$ observations ever published on the BL Lacertae
object GC 0109+224, collected mainly by the robotic telescope of
the Perugia University Observatory in the period November
1994-February 2002. These observations have been complemented by
data from the Torino Observatory, collected in the period July
1995-January 1999, and Mt. Maidanak Observatory (December 2000).
GC 0109+224 showed rapid optical variations and six major
outbursts were observed at the beginning and end of 1996, in fall
1998, at the beginning and at the end of 2000, and at the
beginning of 2002. Fast and large-amplitude drops characterized
its flux behaviour. The $R_c$ magnitude ranged from 13.3 (16.16
mJy) to 16.46 (0.8 mJy), with a mean value of 14.9 (3.38 mJy). In
the periods where we collected multi-filter observations, we
analyzed colour and spectral indexes, and the variability patterns
during some flares. The long-term behaviour seems approximatively
achromatic, but during some isolated outbursts we found evidence
of the typical loop-like hysteresis behaviour, suggesting that
rapid optical variability is dominated by non-thermal cooling of a
single emitting particle population. We performed also a
statistical analysis of the data, through the discrete correlation
function (DCF), the structure function (SF), and the Lomb-Scargle
periodogram, to identify characteristic times scales, from days to
months, in the light curves, and to quantify the mode of
variability. We also include the reconstruction of the historical
light curve and a photometric calibration of comparison stars, to
favour further extensive optical monitoring of this interesting
blazar.
\keywords{BL\ Lacertae\ objects: individual: \object{GC 0109+224}
-- BL\ Lacertae\ objects: general -- quasars: general -- galaxies:
photometry -- methods: statistical}
  }
\maketitle
%

\section{Introduction}
Blazars are an extreme and rare subclass of radio-loud active
galactic nuclei (AGN) showing a broad, non-thermal, polarized, and
highly variable continuum flux, extending over the whole
electromagnetic spectrum. The low-energy emission is likely
synchrotron radiation by relativistic electrons, while the
high-energy one is possibly due to Comptonization of soft photons.
According to the spectral properties displayed, the subgroup of
the classical BL Lacertae objects have been classified into two
categories: the low-frequency peaked (LBL) and the high-frequency
peaked (HBL) (e.g. Padovani \& Giommi 1995; Ulrich et al. 1997). A
population of intermediate BL Lacs, with the synchrotron emission
peaking in the optical band, seems to link the HBL and LBL
classes.

The rapid and violent optical variability is one of the defining
properties of BL Lac objects. Especially the LBL and intermediate
BL Lacs show large-amplitude flares, characterized by flux
variations of even two orders of magnitude and covering a duration
range from hours to years.

Information on variability amplitude, flux variation lags at
different wavelengths, temporal duty cycles, and spectrum changes
can shed light on the location, size, structure, and dynamics of
the non-thermal emitting regions and on the acceleration/radiation
mechanisms. A detailed knowledge of the statistical behaviour on
different time scales is therefore very useful to understand the
basic physical mechanisms in action during the flaring and
quiescent phases. In most cases, optical data collected in the
past are rather sparse and consequently not suitable to perform a
detailed analysis of the blazar variability. International
collaborations and robotic telescopes can improve the amount of
photometric observations by one order of magnitude, so that a
significant sample of these sources can be well observed and
monitored in the optical band with small--medium size dedicated
automatic telescopes.

In this paper we present the results of 1542 optical multiband
photometric observations of the classical BL Lac objetc GC
0109+224 in the $BVR_{c}I_{c}$ Johnson-Cousins filters, during the
period November 1994 -- February 2002. Data were acquired mainly
with the robotic telescope of the Perugia University Observatory,
in Italy, joined by observations performed at the Torino
Observatory, Italy, and at the Mount Maidanak Observatory, in
Uzbekistan. Few Torino observations were published by Villata et
al.\ (1997). Using these observations we analyzed the optical
multiband behaviour of GC 0109+224 and investigated the type of
variability on temporal scales from one day to some months.

The paper is organized as follows: in Sect.\ 2 we briefly review
our knowledge about GC 0109+224. In Sect.\ 3 we describe our
observing and data reduction techniques; calibration of comparison
stars is presented in Sect.\ 4, while our $BVR_{c}I_{c}$ light
curves are discussed in Sect.\ 5. The reconstructed historical
light curve is described in Sect.\ 6; colour indexes are analyzed
in Sect.\ 7, while in Sect.\ 8 we calculate the spectral indexes
describing the variability patterns during distinct flares. In
Sect.\ 9 we perform a statistical analysis of the data, using
well-known tools for unevenly sampled time series. The summary and
conclusion are outlined in Sect.\ 10.
%
%
%

\section{GC 0109+224}

The radio source GC 0109+224, belonging to the Green Bank Radio
Survey List C (other most used names: S2 0109+22, TXS 0109+224, RX
J0112.0+2244, EF B0109+2228, 2E 0109.3+2228, RGB J0112+227), was
first detected in the 5 GHz Survey of the NRAO 43 m disc of Green
Bank, West Virginia (Davis 1971; Pauliny-Toth et al.\ 1972). In
1976 this source was identified with a stellar object of magnitude
15.5 on the Palomar Sky Survey plates by Owen \& Mufson (1977),
who measured a strong millimeter emission (1.53 Jy at 90 GHz) and
defined it as a BL Lac object.

A continuous and featureless optical spectrum was observed, for
the first time, by Wills \& Wills (1979). Pica (1977) recognized
in GC 0109+224 an intermediate optical behaviour between the
large-amplitude variability BL Lacs (e.g.\ BL Lac, ON 231, AO
0235+16) and the smaller-amplitude objects (e.g.\ ON 235, 3C 273).
The host galaxy is unresolved in NTT observations (Falomo 1996)
and UKIRT (K-band) observations (Wright et al.\ 1998). Falomo
(1996) suggested a lower limit to the redshift $z \ge 0.4$ based
on the optical appearance, and assuming $M_{R}=-23.5$ for it, a
value similar to that characterizing some galaxies located at
north-east at some arcseconds from the object. Optical
spectrophotometric observations were compatible with a single
power law (Falomo et al.\ 1994), and there was no evidence for a
thermal component in the far-infrared--optical spectral energy
distribution (Impey \& Neugebauer 1988). Both the optical flux and
the polarization are known to be variable on different timescales,
including intranight variations (Sitko et al.\ 1985; Mead et al.\
1990; Valtaoja et al.\ 1991). The strongly variable degree and
direction of the linear polarization are one of the most
noticeable characteristics of this object (Takalo 1991; Valtaoja
et al.\ 1993). The degree of optical polarization was seen to vary
between 10\% and 30\%, among the highest values observed in this
kind of sources. There is no clear correlation between flux level
and polarization. The frequency-dependence of the polarization
degree is transitory, the position angle is variable and sometimes
frequency-dependent (Valtaoja et al.\ 1993). The observed
near-infrared flux variations are smaller than in the optical one
(Fan 1999).

\begin{figure}[t!]
\begin{center}
\begin{tabular}{c}
{\resizebox{8.5cm}{!}{\includegraphics{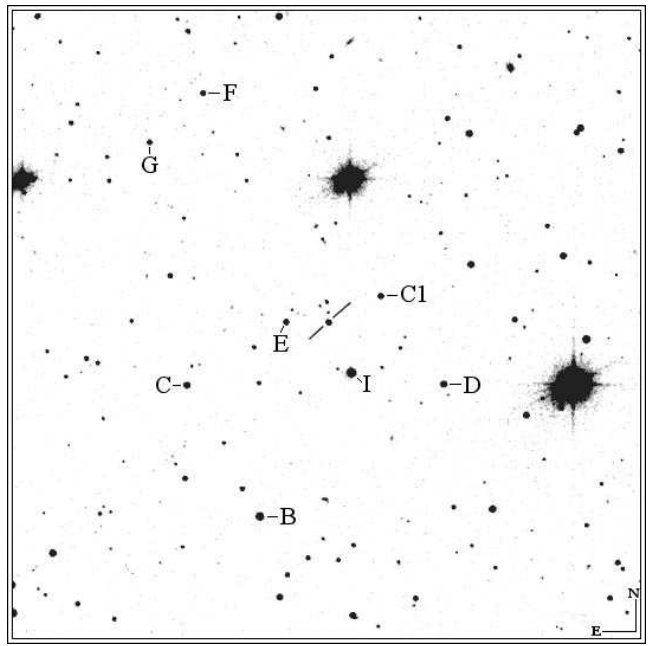}}}\\
\end{tabular}
\end{center}
\vskip -0.5 true cm \caption{Finding chart with a field of
15'$\times$15' centered on GC 0109+224 (elaborated from a frame of
the Digitized Sky Survey). Calibration of stars D, C1, I,  E is
reported in Table \ref{tab:ourcompstars}. Stars B C F G belong to
the photometric sequence calibrated by Miller et al.\ (1983). }
\label{fig:starfield}
\end{figure}
%
%
\begin{table*}[t!]
\caption{$BVR_{c}I_{c}$ Johnson-Cousins photometric calibration of
comparison stars in the field of GC 0109+224.}
\label{tab:ourcompstars} \vspace{-0.5cm}\centering {}
\par
\begin{tabular}{lccccccc}
\vspace{-1mm} \\
\hline \hline
\vspace{-2mm} \\
 star & R.A.& Dec. & $U^{(1)}$ & $B$ & $V$  & $R_{c}$ & $I_{c}$    \\
 &{\footnotesize (J2000.0)} & {\footnotesize (J2000.0)}& {\footnotesize (mag)} & {\footnotesize (mag)} & {\footnotesize (mag)} &  {\footnotesize (mag)}& {\footnotesize (mag)}

   \vspace{1mm} \\
 \hline \hline
\vspace{-3mm} \\
\textbf{D}  &01 11 53.4 &+22 43 17.9  & 15.48 & 15.19 $\pm$ 0.06&
14.45 $\pm$ 0.05 & 14.09 $\pm$ 0.05 & ...
 \\ \hline
\textbf{C1} &01 12 00.3 &+22 45 22.3 & ...  & 16.30 $\pm$ 0.10&
15.28 $\pm$ 0.07& 14.72 $\pm$ 0.06& 14.22 $\pm$ 0.08
  \\  \hline
\textbf{I} &01 12 03.2 &+22 43 30.7 & 13.31 & 13.25 $\pm$ 0.06&
12.51
$\pm$ 0.05 &12.11 $\pm$ 0.04& 11.76 $\pm$ 0.04     \\
\hline
 \textbf{E}  &01 12 10.6 &+22 44 40.3 & 15.78 & 16.01 $\pm$ 0.08& 15.29 $\pm$ 0.07 & 14.94 $\pm$ 0.05 & 14.60 $\pm$ 0.07
 \\
 \hline \hline
\end{tabular}
\begin{list}{}{}
\item[$^{\mathrm{(1)}}$] $U$ values by Miller et al.\ (1983)
\end{list}

\end{table*}
%
%

In the radio bands the source is variable in flux, degree of
polarization and position angle, showing a flat average spectrum
as expected for classical BL Lacs. At the VLA milliarcsecond (pc)
scale the 5 GHz radio map of GC 0109+224 reveals a compact core
with a secondary component at about 3 mas at a position angle
$\simeq85^{\circ}$, with no important additional diffuse emission,
and less luminous and/or beamed than the 1-Jy sample of BL Lacs
(Bondi et al.\ 2001). The same milliarcsecond structure was
evident in VLBA images at 2 and 8 GHz (Fey \& Charlot 2000). The
kpc scale shows a faint one-sided collimated radio jet, about 2
arcsec long, in south-western direction (Wilkinson et al.\ 1998),
largely misaligned with the pc-scale inner region. This
misalignment between structures at pc and kpc scales has
frequently been observed in high-luminosity LBL. The 200-mJy
sample of blazars (March\~{a} et al.\ 1996), including GC
0109+224, seems to fill the gap between HBL and LBL, as expected
by unification pictures (Bondi et al.\ 2001). In particular, GC
0109+224 is apparently placed on the boundary between the LBL and
intermediate obects. This source is regularly monitored by the
University of Michigan Radio Astronomy Observatory (UMRAO), in
USA, and by the Mets\"ahovi Radio Observatory, in Finland.\par It
was observed in the X-ray band by HEAO-1 (Della Ceca et al.\
1990), \textit{Einstein} Observatory (HEAO-2, Owen et al.\ 1981),
EXOSAT (Maraschi \& Maccagni 1988, Giommi et al.\ 1990; Reynolds
et al.\ 1999), and ROSAT (Neumann et al.\ 1994; Brinkmann et al.\
1995; Kock et al. 1996, Reich et al.\ 2000). The source is also a
member of the RGB (ROSAT All-Sky Survey-Green Bank) catalog by
Laurent-Muehleisen et al.\ (1999), a uniform survey of classical
BL Lacs generated from cross-correlating the 6 cm deep radio
images with the X-ray samples. It consists in a sample of
intermediate BL Lacs, with properties (like the fluxes ratio
$F_{X}/F_{r}$) smoothly distributed in a large range between the
LBL and HBL subclasses: for example, in the radio-optical versus
optical-X spectral indexes plot, which approximately divides the
HBL and LBL populations, GC 0109+224 appears close to a typical
intermediate object like ON 231 (Dennett-Thorpe \& March\~{a}
2000).

GC 0109+224 was not detected in $\gamma$-rays by EGRET (Fichtel et
al.\ 1994), with a rather low upper limit of 0.01 nJy (above 100
MeV).

This source has recently been included in the list of blazars
chosen for a project of continuous optical monitoring named the
Whole Year Blazar Telescope (WYBT; Tosti et al.\ 2002), born from
the wide international collaboration called Whole Earth Blazar
Telescope
(WEBT\footnote{\texttt{http://www.to.astro.it/blazars/webt/}};
Villata et al.\ 2000, 2002).
%
%

\section{Observations and data reduction}
The photometric observations were carried out with three
instruments: the Newtonian f/5, 0.4 m, Automatic Imaging Telescope
(AIT) of the Perugia University
Observatory\footnote{\texttt{http://wwwospg.pg.infn.it}}, Italy
(451 meters a.s.l.), a robotic telescope equipped with a $192
\times 165$ pixels CCD array, thermoelectrically cooled with
Peltier elements (Tosti et al.\ 1996); the REOSC f/10, 1.05 m,
astrometric reflector of the Torino Observatory, Italy (622 meters
a.s.l.), mounting a $1242 \times 1152$ pixel CCD array, cooled
with liquid nitrogen and giving an image scale of $0.467''$ per
pixel; the Ritchey-Chr\'{e}tien f/7.74, 1.5 m, AZT-22 telescope
(Novikov 1987) of the Mount Maidanak Observatory, Uzbekistan (2593
meter a.s.l.), equipped with a nitrogen cooled SITe $2048 \times
800$ pixel CCD device, with a $8.5\times 3.5'$ field of view.\par

The telescopes were provided with standard $UBV$
(Johnson) and $R_{c}I_{c}$ (Cousins) filters (Bessel 1979).
\begin{figure*}[t]
\centering
\resizebox{\hsize}{!}{\includegraphics[angle=0]{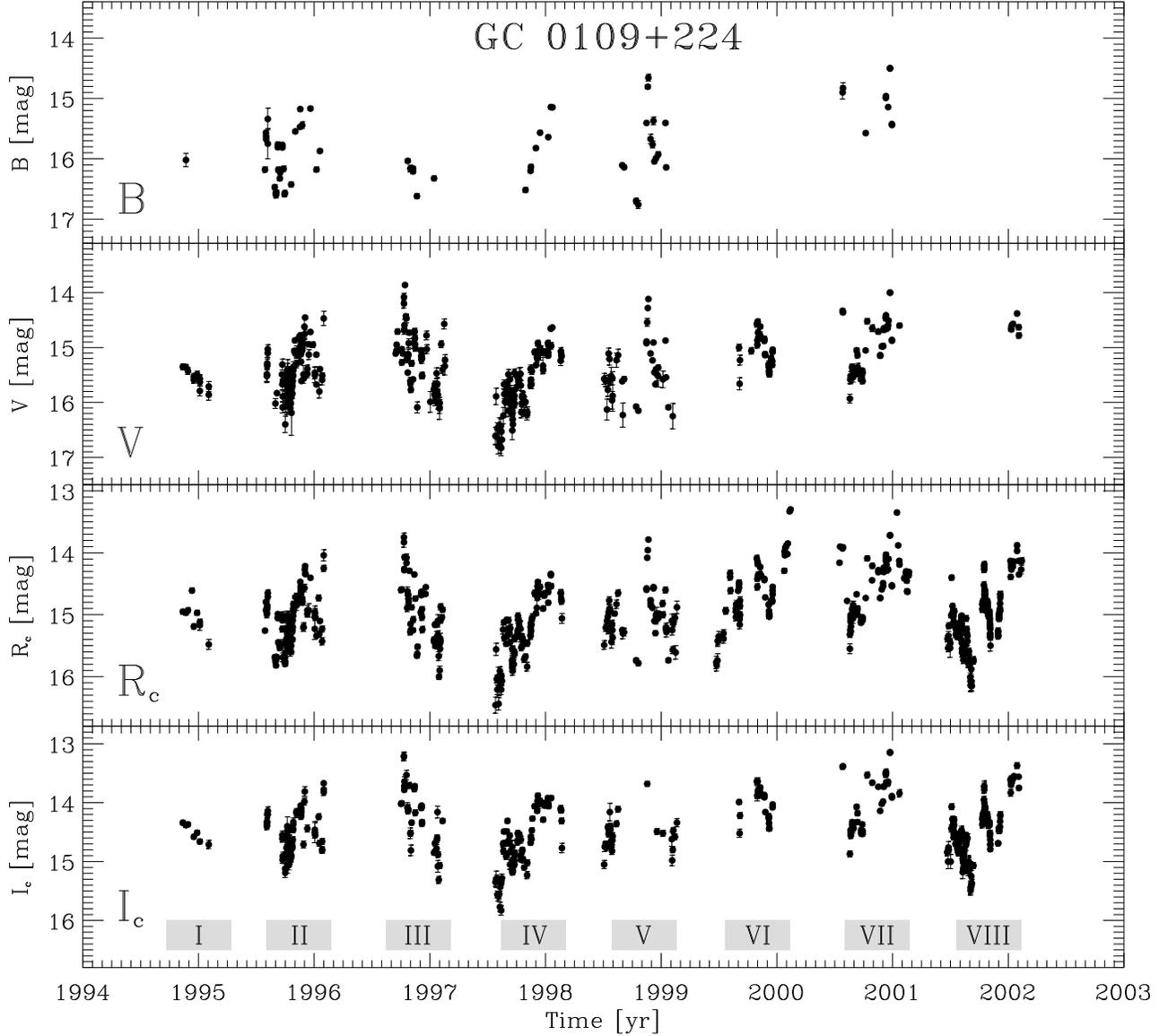}}
\vspace{-0.5 cm} \caption{$BVR_{c}I_{c}$ light curves of GC
0109+224 from 1994 to 2002. All the data came from our seven-years
observing campaign. The roman numbers point out the number of each
observing season.} \label{fig:clucedatinostri}
\end{figure*}
All observatories took CCD frames and performed a first automatic
data reduction with batch procedures, using standard methods, to
correct each raw image for dark and bias signals and to flat
fielding, to recognize the field stars, and to derive instrumental
magnitudes via aperture photometry or Gaussian fitting. The single
frames are inspected to evaluate the quality of the image, the
reliability of the data, and to search for spurious interferences.
The comparison among data obtained with different telescopes in
the same night reveals a good agreement, and no revealable offset
was found, as already demonstrated for another source (Raiteri et
al. 2001). A comparison with data in $V$ band taken in the same
period at the Tuorla Observatory (Katajainen et al.\ 2000)
revealed a good agreement.

\section{Comparison stars photometry}
Calibration of the source magnitude is easily obtained by
differential photometry with respect to comparison stars in the
same field. In order to obtain a reliable photometric sequence for
GC 0109+224 we selected a set of non-variable stars with
brightness comparable to the object and different colours (see the
finding chart in Fig.~\ref{fig:starfield}). Calibrations of these
stars was derived from several photometric nights at the Perugia
and Torino Observatories using Landolt standards. Details on
observing and data reduction procedures, filter system, and
software adopted, as well as a comparison with other works can be
found in Fiorucci \& Tosti (1996), Fiorucci et al.\ (1998),
Villata et al.\ (1998), and Raiteri et al.\ (1998). Data obtained
by the two telescopes showed an agreement into one standard
deviation, and a weighted mean on the nights number was finally
adopted. The results are presented in
Table~\ref{tab:ourcompstars}, showing our $B,V,R_{c},I_{c}$
photometry as well as the $U$ magnitudes reported by Miller et
al.\ (1983).

\section{Optical light curves from 1994 to 2002}
We have been monitoring the BL Lac object GC 0109+224 in the four
$B,V,R_{c},I_{c}$ optical bands for more than 7 years, from
November 12, 1994 to February 12, 2002 (JD=2449669--2452318). A
total of 1542 data points was obtained in a period of 2649 days
(see Fig.\ \ref{fig:clucedatinostri}). In particular, we collected
671 photometric points in the $R_{c}$ band (the best sampled one)
during eight observing seasons: the mean time lag between
subsequent observations is $\Delta t = 3.9$ days, the minimum lag
$\Delta t = 6$ minutes and the maximum lag $\Delta t = 244$ days,
corresponding to the interruption between the second and the third
observing period.

From a direct visual inspection of the light curves structure in
Fig.\ \ref{fig:clucedatinostri}, we can identify an intermittent
mode of variability, a common behaviour among BL Lacs (Tosti et
al. 2001). Several moderate-amplitude outbursts are visible,
together with some steep brightness drops and some periods of
flickering at an intermediate level. No extraordinary big and
isolated flare was detected, but six major flares are evident: at
the beginning and fall of 1996, in fall 1998, at the beginning and
at the end of 2000, and at the beginning of 2002.

In the third observing season (October 1996 -- February 1997), an
appreciable flare was detected between October and November,
peaked on JD=2450367 (Oct.\ 11, 1996), when GC 0109+224 reached
the magnitudes $V=14.08\pm 0.07$ and $R_{c}=13.75\pm 0.07$. The
variability pattern of this flare is inspected in Sect.\ 8. The
flux then dropped by almost two mag in 42 days (at JD=2450409
$R_{c}=15.63\pm 0.07$), and began to fluctuate around a low level.
In the fourth observing season (July 1997 -- February 1998) we
observed a quasi-monotonic ascendant trend with small fluctuations
on different time scales superimposed. At the end of this climbing
phase the $R_{c}$ magnitude has passed from the minimum brightness
value detected by us ($16.46\pm 0.13$) to $14.341\pm 0.025$. In
the period October--December 1998 (in the fifth observing season),
around JD=2451139 (Nov.\ 21, 1998) the source reached magnitudes
$B=14.657\pm 0.056$ and $R_{c}=13.786 \pm 0.026$, according to
Torino and Perugia $BVR_{c}$ observations. A similar behaviour is
observed in the sixth season (June 1999 -- February 2000), where
the brightness rises from an initial value of $R_{c}=15.81\pm 0.1$
at JD=2451353 to $R_{c}=13.30\pm 0.03$ (the maximum brightness
observed in this band), showing at the end a prominent flare. In
November--December 2000 another flare (peaking at JD=2451902
according the $B$ mag) is clearly observed by both the Mt.\
Maidanak and the Perugia Observatories in all $UBVR_{c}I_{c}$
filters. The variability pattern is analyzed also in this case, in
detail in Sect.\ 8.
\begin{figure*}[t!] \centering
\resizebox{\hsize}{!}{\includegraphics[angle=90]{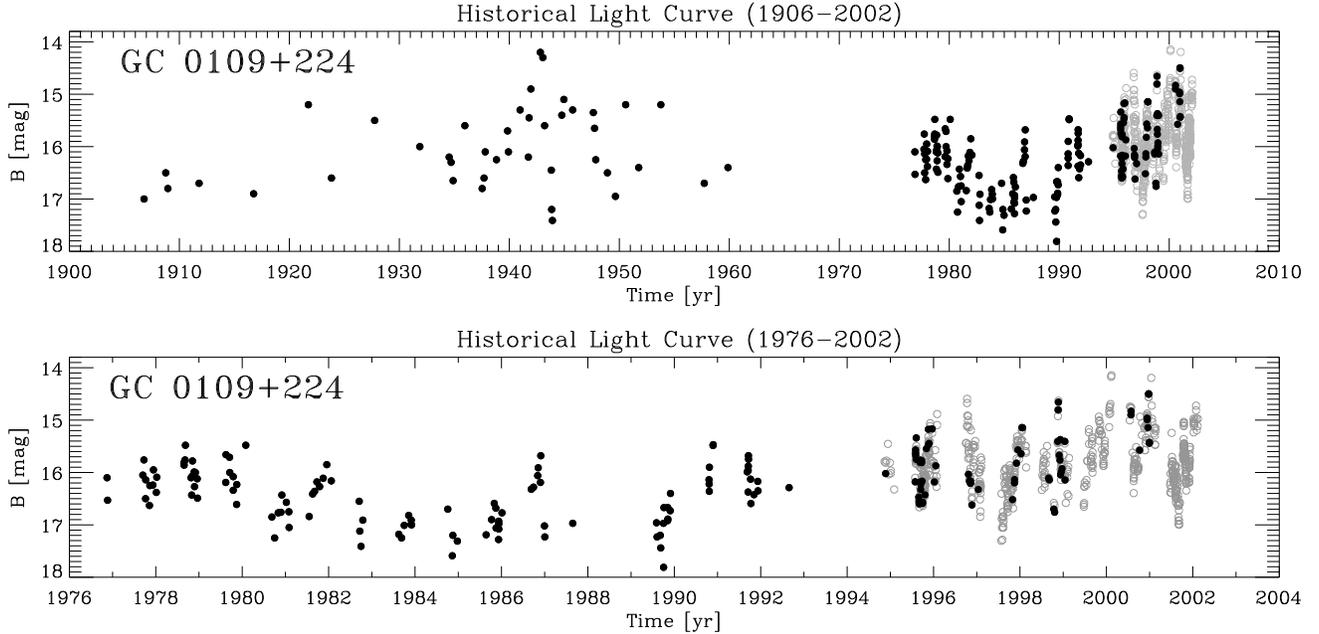}}
\vspace{-0.2cm} \caption{The historical light curve of GC 0109+224
in $B$ band (black dots) reconstructed from the literature (Pica
1977; Pica et al.\ 1980; Puschell \& Stein 1980;  Zekl et al.\
1981; Moles et al.\ 1985; Pica et al.\ 1988; Xie et al.\ 1988a;
Xie et al.\ 1988b; Mead et al. 1990;
Sillanp$\ddot{\textrm{a}}\ddot{\textrm{a}}$ et al.\ 1991; Takalo
1991; Xie et al.\ 1992; Valtaoja et al.\ 1993; Falomo et al.\
1994; Xie et al.\ 1994; Katajainen et al.\ 2000). Data after 1994
are almost entirely from this paper. $B$ magnitudes derived from
$R_c$ values by using the mean $B-R_c$ index are added to improve
sampling (grey circles); the maximum error for these estimates is
about 0.3--0.4 mag. Pre-1960 values have original errors of 0.1
mag and represent photographic magnitudes (see text). Error bars
are not represented for clarity.} \label{fig:storica}
\end{figure*}

The fainter and brighter magnitudes measured in the $B$ band are
$16.76\pm 0.06$ (at JD=2451107, Oct 20 1998) and $14.497\pm 0.005$
(JD=2451902, Dec 23 2000) respectively, while the fainter and
brighter magnitudes observed in the better sampled $R_{c}$ band
are $16.46\pm 0.13$ (at JD=2450656, Jul 27 1997) and $13.30\pm
0.03$ (at JD=2451588, Feb. 13, 2000) respectively.

A general visual inspection of the light curves in Figs.\ 2-3 reveals
the existence of a long-term oscillation of the base flux level, from which
rapid flares depart.

\section{The historical light curve}
The optical history of GC 0109+224 extends over almost one century
(1906--2002), even if there is a gap of 15 years (1960--1975) in
the available data (Fig. \ref{fig:storica}). The old light curve
was obtained using plates, from which a photographic magnitude
$m_{pg}$ can be extracted. Recent data have been obtained directly
with photoelectric or CCD photometry. A semi-empirical correction
must be applied to convert $m_{pg}$ data in $B$ magnitudes. For
example, one can adopt the mean shift $B=m_{pg}+0.28$, as
suggested by L\"{u} (1972), but the exact value is doubtful, due
to its dependence on the variable $U-B$ index. For this reason, we
chose to report the original non-corrected data, warning that
pre-1961 values are overestimated in luminosity by $\sim 0.2$--0.3
mag. The optical behaviour before 1961 was reconstructed with the
Heidelberg plates (Zekl et al.\ 1981) and the Harvard plate
collection (Pica 1977). The light curve between 1976 and 1988 was
mainly obtained at the University of Florida Rosemary Hill
Observatory (Pica et al.\ 1988). The other observations are sparse
literature data (see references in Fig. \ref{fig:storica}). Data
after 1994 are almost entirely from this paper. To give a complete
qualitative behaviour, we have reported in Fig.\
\ref{fig:storica}, in addition to the original $B$ magnitudes,
also rough estimates of the $B$ values derived from our $R_{c}$
mag, using the mean $B-R_{c}$ color index, which is approximately
constant in our period of observation (see Sect.\ 7).

The largest brightness variation occurred in 1942--1943, when a
steep decline of 3.07 mag during one year was observed (Pica
1977), following one of the maximum values ever achieved
($m_{pg}\simeq14.2$). A comparable brightness level was reached
again only in the flares of October 1996, of November 1998, in
February 2000 and in the flare of December 2000.

We also notice that from 1944 to 1996  the source was never
observed brighter than $B=15$, even if this could be due to poor
sampling. Variations of 0.6--0.8 mag and modest flaring were
common from 1976 to 1995. In 1980 a drop of almost 2 mag made the
source fainter than $B=17$, and in the period 1981--1990 the
object remained in a low state around $B=17$ except for the flares
of late 1981 and 1986. In particular, in August 1989 the source
brightness fell down to $B=18.42$, the minimum value ever observed
(Takalo 1991). In the post-1994 better sampled light curve
obtained with our data, we see a general brighter mean $B$ value
(15.8), again comparable with the 1920--1960 period, and some
flare events. Variations of $\sim$2.5 mag in less than one year
are common (a drop of 2.25 mag in 1996--97, a brightening of 2.1
mag in 1997--98 and of 2.51 mag in 1999--2000, a drop of 2.8 mag
in 2001 and finally an increase of 2 mag in 2001--2002).

\begin{figure}[t!]
\centering
\begin{tabular}{c}
\hspace{-0.5cm}
{\resizebox{\hsize}{!}{\includegraphics{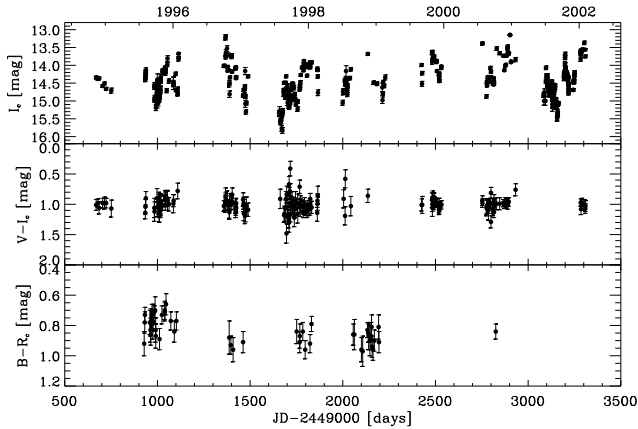}}} \\
\end{tabular}
\caption{Temporal behaviour of the $I_{c}$ magnitude (upper panel)
and of $V$-$I_{c}$ (middle panel) and $B$-$R_{c}$ (lower panel)
colour indexes.} \label{fig:indexcol2}\vspace{-0.5cm}
\end{figure}

\section{Colour indexes}\label{par:colorindexes}
Optical flux variations in BL Lacs are usually accompanied by
changes in the spectral shape, and this can be revealed by
analyzing color indexes. We calculated the $B-R_{c}$ and $V-I_{c}$
indexes, selecting data from the same observatory, coupling frames
with a lag of no more than 15 minutes, and using only the most
precise data. Since the GC 0109+224 host galaxy is faint it is
reasonable to neglect the galaxy color interference in the
observed flux, which instead may be important in other objects
(like BL Lacertae, see e.g.\ Villata et al.\ 2002).

When plotting these two color indexes versus magnitude in
different bands, we always found a certain dispersion of data
around the mean values ($B-R_{c}=0.83 \pm 0.08$ and $V-I_{c}=1.00
\pm 0.16$). The $B-R_{c}$ index is a good indicator of reddening
variations during the different variability phases, but neither
solid correlations, characterized by a small dispersion of the
data, nor general trends seem to be present. This means that the
overall behaviour of the indexes does not depend on the source
brightness.

Fig.\ \ref{fig:indexcol2} reports the long-term temporal evolution
of the $V-I_{c}$ index, which appears to spread around the mean
value without any noticeable trend. A larger data dispersion with
larger data errors occurs when the source is faint. The less
sampled $B-R_{c}$ temporal trend shows a small but clear and
abrupt increase in 1996, from the mean value of 0.78 in July 1995
to the mean value of 0.94 in November 1996, meaning a spectral
steppening. In the following years this color index remains nearly
constant.

In the overall datasets, we found the same no-trend spreading
when calculating the other color indexes and plotting them versus
magnitudes. In the flares temporal windows the situation is
different: a chromatic behaviour appears evident, as showed in the
next section for three examples.

\begin{table}[t!]
\caption[]{The number of photometric $U B V R_{c} I_{c}$ data
points of GC 0109+224 obtained by each observatory, and a summary
of the data sampling and flux in each band.}
\label{tab:samplingprop} \vspace{-0.2cm} \centering {}
\begin{tabular}{ccc}
\vspace{-3mm} \\
\hline \hline
$~~~~~~~~~$   & \scriptsize{NUMBER OF DATA PER OBSERVATORY} & $~~~~~~~~~$   \\
\end{tabular}
\begin{tabular}{lccccccc}
\vspace{-4mm} \\
\hline
\vspace{-3mm} \\
 Obs.& $U$ & $B$  & $V$& $R_{c}$ & $I_{c}$ & Tot. & \scriptsize{Period}  \\
\hline \hline
Perugia &   0 &  5 &  309 & 568 & 434 & 1316 & \scriptsize{Nov94-Feb02}\\
Torino  &    0  & 56 & 41 & 96 & 0  & 193    & \scriptsize{Jul95-Jan99}\\
Maid.&    5 &  7 &  7  & 7 &  7 &  33        & \scriptsize{Dec00}\\
\hline
Total & 5  & 68 & 357 & 671 & 441 & 1542 \\
\hline \hline
\end{tabular}
$~$\\ $~$ \\
\begin{tabular}{ccc}
\vspace{-3mm} \\
\hline \hline
 $~~~~~~~~~~~~~~~~~~~$   & \scriptsize{SAMPLING AND FLUXES} & $~~~~~~~~~~~~~~~~~~$   \\
\end{tabular}
\begin{tabular}{lcccc}
\vspace{-4mm} \\
\hline
\vspace{-3mm} \\
$~$ &  $B$  & $V$& $R_{c}$ & $I_{c}$   $~$  \\
\hline \hline
$~$Start date [JD-2449000]&     679 & 669& 669& 669$~$   \\
$~$End date [JD-2449000]&     2908& 3309 &  3318 & 3309$~$   \\
$~$Mean gap in data [day]&  33.2  & 6.7 & 3.9 & 6.0$~$   \\
$~$Longest time gap [day]&       557 & 352 & 244 & 244$~$   \\
$~$Mean flux [mJy] &  2.39      & 2.88 & 3.38 & 4.47$~$   \\
$~$Max flare flux [mJy] & 7.85  & 11.66  & 16.16 & 15.10$~$   \\
$~$Max/Min flux ratio & 6.87  & 10.90 & 15.28 &  14.63$~$  \\
$~$Absorption coeff. [mag] & 0.161 & 0.124 & 0.100 & 0.073$~$  \\
\hline \hline
\end{tabular}
\end{table}

\section{Spectral indexes and variability patterns}

In this section we examine possible relations between the source
luminosity and the shape of the optical spectral energy
distribution (SED), which can be expressed conveniently by a power
law $\nu F_\nu \propto \nu^{-\alpha+1}$, ($\nu$ being the
radiation frequency and $\alpha$ the spectral index). We have
already partially investigated this with the analysis of the color
indexes.

The degree of correlation between $\alpha$ and the flux
sheds light on the non-thermal emitting processes, involving
synchrotron and inverse-Compton cooling of a population of
relativistic electrons. A common pattern, but not the unique one,
during well-defined flares is the soft-hard-soft
signature: the brighter the source, the harder the spectrum, in
the sense that the spectral slope flattens when the source
luminosity increases. This may suggest that the more intense is the
energy release, the higher is the particles energy. The
non-thermal acceleration mechanism tends to work with almost
fixed populations of particles, in confined flaring regions in
the jet.

Using data of a single telescope in at least three filters
with maximum time lag of 15 minutes, we checked the degree of
correlation between $\alpha$ and the flux in various bands through
a least-square linear regression. Values characterized by large
errors and $\chi^2$ were rejected. We found that the spectral
index of  GC 0109+224 varies between $2.23$ to $0.82$, with a mean
value of $1.41\pm 0.16$.

\begin{figure}[t!]
\centering
\begin{tabular}{l}
\hspace{-5mm}
{\resizebox{\hsize}{!}{\includegraphics{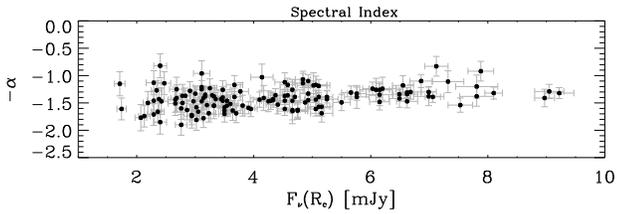}}} \\
\end{tabular}\vspace{-2mm}
\caption{Spectral index $\alpha$ versus
flux in the $R_c$ band. The linear correlation coefficient is
$r_{\alpha-R}=0.29$. A weak indication of spectral
flattening with brightness is recognizable (slope $0.039$).}
\label{fig:alphavsflux}
\end{figure}
\begin{figure}[t!]
\centering
\begin{tabular}{l}
\hspace{-5mm}
{\resizebox{8cm}{!}{\includegraphics{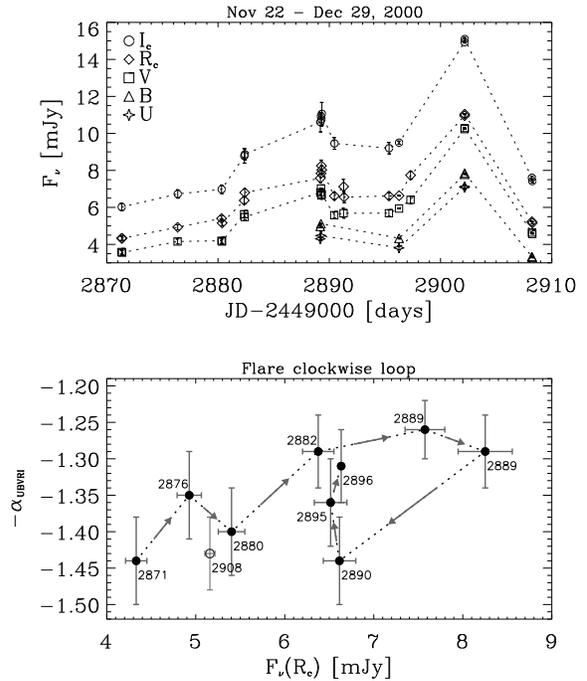}}} \\
\end{tabular}
\caption{Evolution of the optical spectrum of GC 0109+224 as a
function of the flux in the $R_c$ band during the double-peaked
flare of November 22 -- December 29, 2000. The loop formed by
points connected by arrows corresponds to the first peak. In the
subsequent major peak the flux achieves $F_{\nu}(R_c)=11.04$ mJy
and the spectral index $\alpha=1.09$ on day 2902 (this point is
out of the plot, for clarity), and then drops rapidly to the base
value $F_{\nu}(R_c)=5.15$ mJy, $\alpha=1.39$, on day 2908 (empty
circle in the plot). } \label{fig:loop00}
\end{figure}
\begin{figure}[t!]
\centering
\begin{tabular}{l}
\hspace{-5mm}
{\resizebox{8cm}{!}{\includegraphics{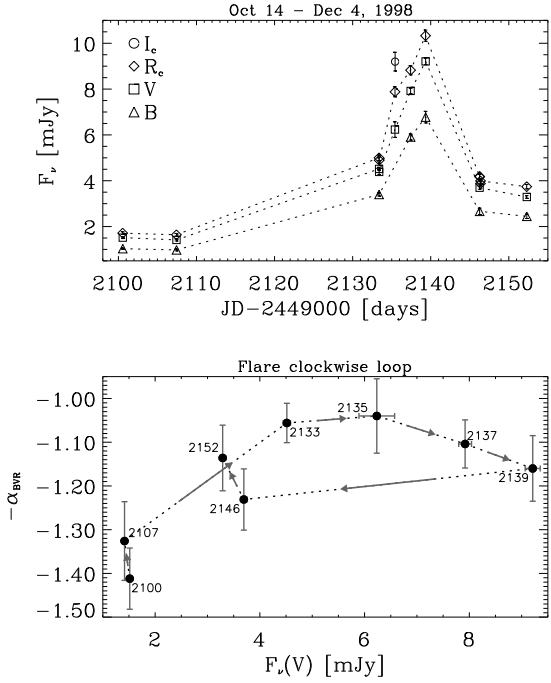}}} \\
\end{tabular}
\caption{Evolution of the optical spectrum of GC 0109+224 as a
function of the flux in the $V$ band during the isolated flare of
14 October -- 4 December 1998. A well-defined clockwise loop is
traced evidencing, one more time, that the spectrum gets softer
when the source gets fainter.} \label{fig:loop98}
\end{figure}
In Fig.\ \ref{fig:alphavsflux} we plotted $\alpha$ versus
$R_c$ flux. We found correlation for the 141 data points of
the Perugia Observatory, where the linear correlation coefficients for the
different bands are $r_{\alpha-I}=0.18$ (pure chance probability
less than 0.04), $r_{\alpha-R}=0.29$, $r_{\alpha-V}=0.36$ (chance
probability 0.01), and in the 18 Torino Observatory data points,
with the value $r_{\alpha-B}=0.60$ (chance probability 0.008).
This increasing trend of $r$ with frequency derives
from the stronger variability at higher frequency, which in turn determines the
``flatter when brighter" behaviour. However, also uncorrelated random
fluctuations in the emission might introduce a statistical bias
due to the spectral index dependency on the flux (Massaro \&
Trevese 1996). An unbiased estimate of the correlation coefficient
is given by the value computed for the central frequency (in this
case close to that of the $R_{c}$ band), which we used as
representative of the brightness state of the source. In Fig.\
\ref{fig:alphavsflux} one can also notice a weak indication of spectral
flattening with increasing brightness,
characterized by a slope of $0.039$.

During well-defined and large flares observed in blazars at the
X-ray frequencies, the spectral index versus flux shows a
characteristic loop-like pattern in most cases. This is usually
tracked temporally in the clockwise sense (if the power index is
plotted entirely with sign). This feature is well known and has
been observed for example in OJ 287 (Gear et al.\ 1986), PKS
2155-304 (Sembay et al.\ 1993; Georganopoulos \& Marscher 1998;
Kataoka et al.\ 2000), Mkn 421 (Takahashi et al.\ 1996), and H
0323+022 (Kohmura et al.\ 1994). The loops represent an hysteresis
cycle in the scatter plot between the power index and the flux,
and indicate that the spectrum gets steeper when the source gets
fainter. They arise whenever the spectral slope is completely
controlled by the radiative cooling processes, so that the
information about changes in the injection rate of accelerated
particles propagates from high to low energies (Kirk et al.\ 1998;
Kirk \& Mastichiadis 1999). In particular, clockwise loops mean
that cooling is effective before acceleration has ceased.

In order to check the presence of this behaviour in our optical
data, we selected three well-defined and strong flares of GC
0109+224, lasting less than one month. We discovered a clockwise
loop from the double-peaked flare of November 22 -- December 29,
2000. Fig.\ \ref{fig:loop00} shows that the cycle formed by the
first peak is well traced by our data. From the subsequent major
peak we obtained only two points, corresponding to the maximum
brightness (and to the flattest spectral index) and to the
minimum one, where the flux comes back to the base level from
which the first peak started.

A well-defined
clockwise loop comes out from the isolated flare of October 14 --
December 4, 1998 (Fig.\ \ref{fig:loop98}).
The same signature of a steeper-when-dimming trend is clear.

\begin{figure}
\centering
\begin{tabular}{l}
\hspace{-5mm}
{\resizebox{8cm}{!}{\includegraphics{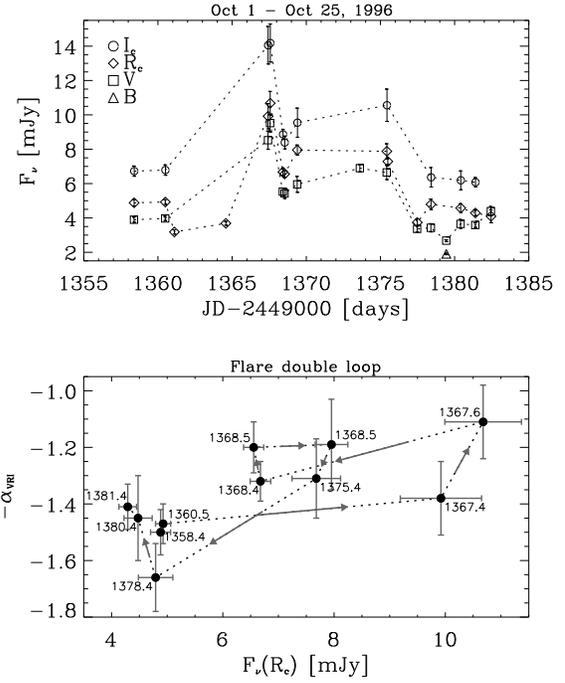}}} \\
\end{tabular}
\caption{Evolution of the optical spectrum of GC 0109+224 as a
function of the flux in the $R_c$ band during the double-peaked
flare of October 1--25, 1996. An anticlockwise loop due to the
first peak is superimposed to a second clockwise loop due to the
second peak. This might be a general and nice signature of flares
superimposition.} \label{fig:loop96}
\end{figure}

The plot obtained for the double-peaked flare lasting from October
1, 1996 to October 24, 1998 is displayed in Fig.\
\ref{fig:loop96}. It shows an anticlockwise loop due to the first
peak superimposed to a second clockwise loop due to the second
peak. This could be a nice signature of the superimposition of a
second flare on a previous one. The errors on $\alpha$ are large,
but the data are nonetheless consistent with the discussed pattern,
because of the smaller uncertainty on the flux. Anticlockwise
loops are not common but have occasionally been observed (for
example in the case of PKS 2155-304, Sembay et al.\ 1993),
attesting in this case, that cooling times are comparable to
acceleration times.

Our present findings show that variability loops,
during isolated flares that do not blend with any preceding or
subsequent variability bump, are recognizable not only in the
X-ray emission, but also in the optical one, once
sampling is sufficient to trace the patterns well.
This suggests that during flares of the kind described above, extending over
few weeks, radiative cooling dominates the spectral distribution also at the
optical frequencies.

Consequently, the variations at higher frequencies (for example in
the $U$ and $B$ bands) lead those at the lower frequencies (for
example $R_{c}$ and $I_{c}$ bands) during both the increasing and
decreasing brightness phases, reflecting the differences
in the electron cooling times.

\section{Statistical analysis of the time scales}
In order to investigate the nature and the temporal structure of
variability, the auto-correlations, the existence of
characteristic time scales and possible periodicity, we applied
three well-tested methods: the structure function (SF), the
discrete correlation function (DCF), and the discrete Fourier
transform in the Lomb-Scargle implementation (periodogram). These
methods, optimized for unevenly sampled datasets, give a
quantitative statistical description of the time variability.

The SF provides information on the time structure of a data train
and it is able to discern the range of the characteristic time
scales that contribute to the fluctuations (see e.g.\ Rutman 1978;
Simonetti et al.\ 1985; Neugebauer et al.\ 1989; Hughes et al.\
1992; Smith et al.\ 1993; Lainela \& Valtaoja 1993; Heidt \&
Wagner 1996; Paltani et al.\ 1997, Paltani 1999). It represents a
measure of the mean squared of the flux differences
$(a_{i}-a_{i+\Delta t})$  of $N$ pairs with the same time
separation $\Delta t$ ($a_{i}$ is the discrete signal at time
$t$). We use only the first order auto-SF, defined as
\begin{equation}\label{eq:udcf}
 {\rm SF}^{(1)}(\Delta t)=\frac{1}{N}\sum_{i=1}^{N}\left(a_{i}-a_{i+\Delta
 t}\right)^{2}.
\end{equation}
The general definition involves an ensemble average. For a
stationary random process, the SF is related to the variance
$\sigma^{2}$ and the autocorrelation function ${\rm ACF}(\Delta
t)$ by ${\rm SF}^{(1)}(\Delta t)=2\sigma^{2}\left[1-{\rm
ACF}(\Delta t)\right]$.

The SF is equivalent to the power spectrum, with the advantage to
work in the time domain, which makes it less dependent on
sampling. It removes any continue component (mean value in a
period, or direct current offset) from a signal, while SF of order
$n$ would remove polynomials of order $n-1$. Deep drops in the SF
mean a little variance and then the signature of possible
characteristic time scales.

Typically, the SF increases with $\Delta t$ in a log versus log
representation, showing, in the ideal case, an initial plateau for
short time lags, and a second plateau for lags longer than the
maximum correlation time scale. A steep curve, whose slope depends
on the fluctuations nature, links these two regions.
%
\begin{figure}[t!]
\begin{center}
\begin{tabular}{c}
{\resizebox{7.2cm}{!}{\includegraphics{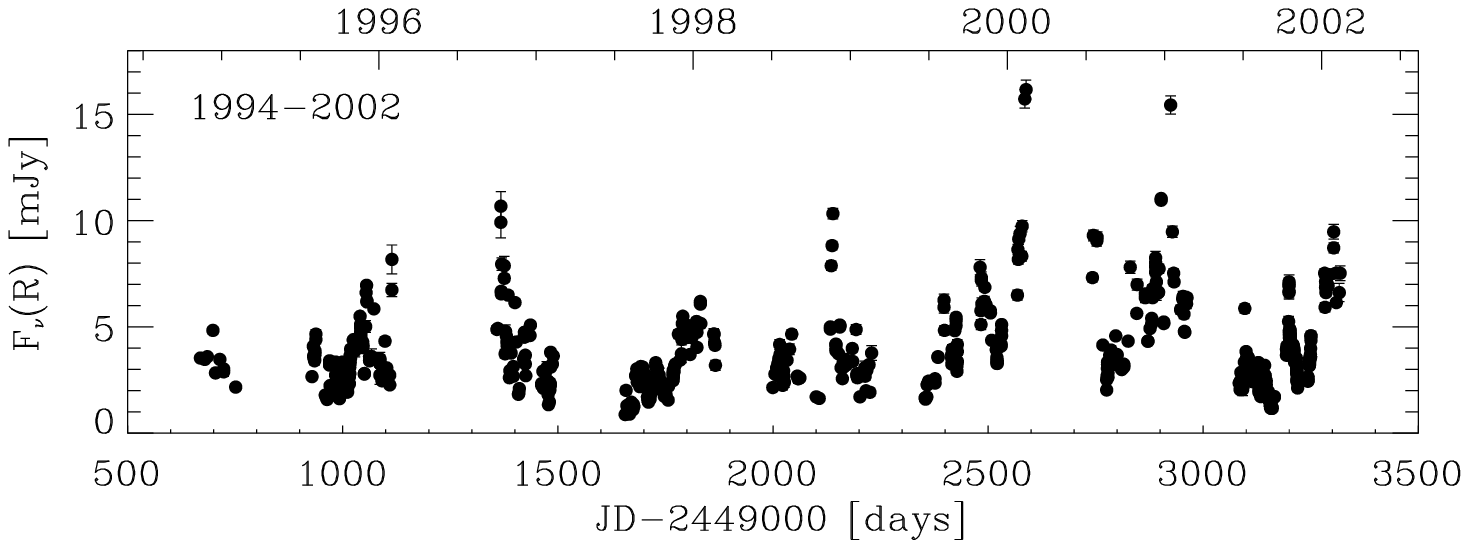}}}\\[-3mm]
{\resizebox{7.2cm}{!}{\includegraphics{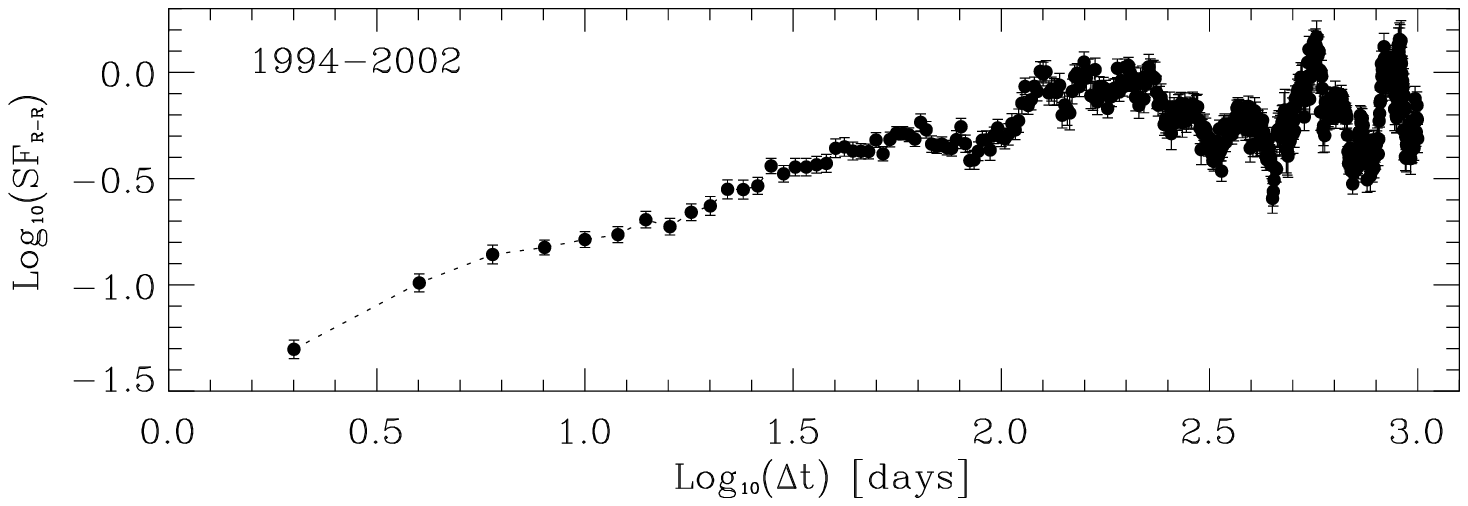}}} \\[-3mm]
{\resizebox{7.2cm}{!}{\includegraphics{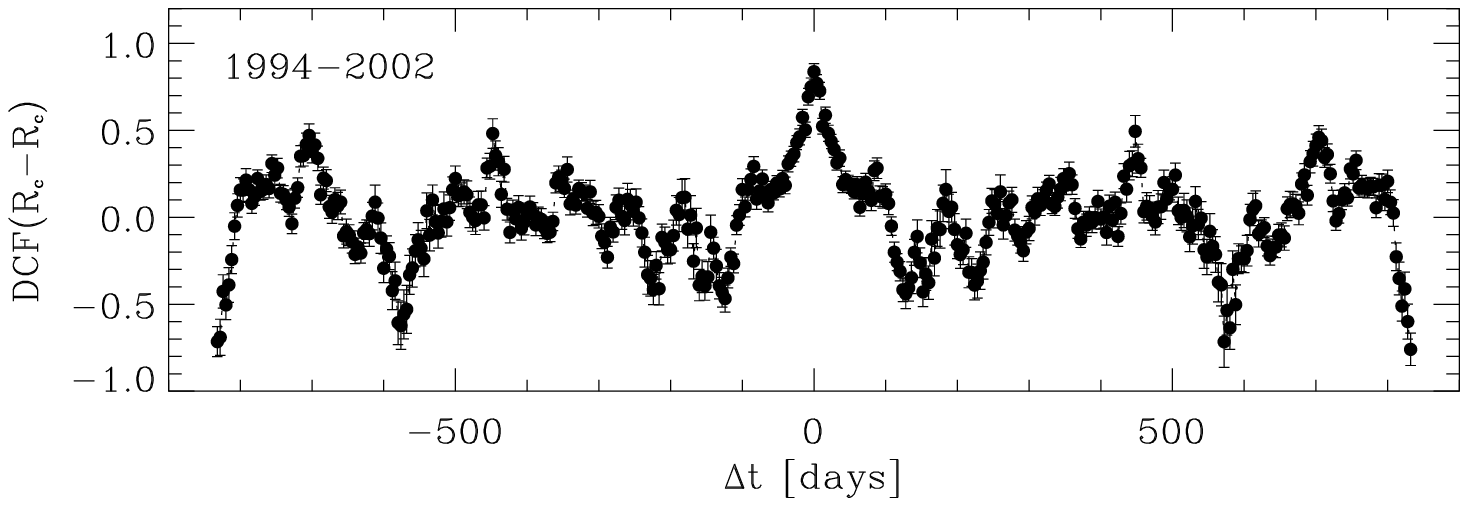}}} \\[-3mm]
{\resizebox{7.2cm}{!}{\includegraphics{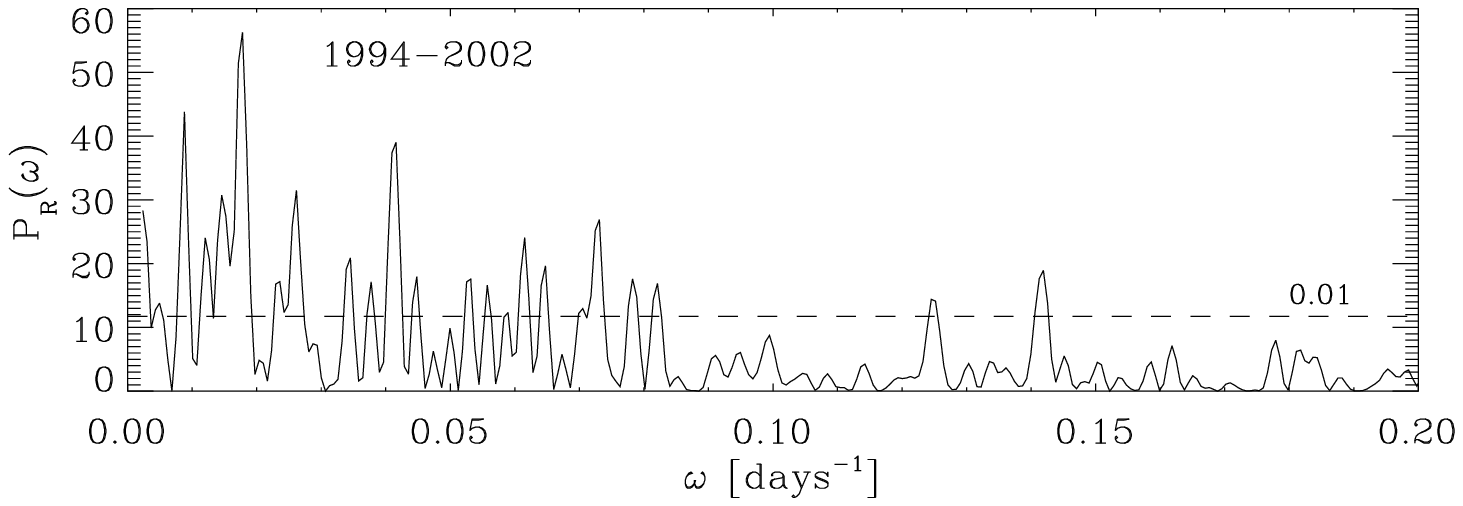}}} \vspace{-5mm}
\end{tabular}
\end{center}
\caption{The 1994--2002 $R_{c}$ flux light curve, its structure
function (data bin: 1 day, SF bin: 2 days), discrete correlation
function (data bin: 1 day, DCF bin: 4 days), and periodogram
($\omega=2\pi f$). The dashed line indicates the 99\% significance threshold.}
 \label{fig:sf-dcf94-02}
\end{figure}
The turnover time lag between the rising part and the long-lag
upper flattening of the SF identifies a characteristic variability
time scale of the source (Hughes et al.\ 1992; Lainela \& Valtaoja
1993). The slope $\beta$ of the intermediate part is related to
the index of the power spectral density function (PSD). A
``typical'' PSD has a power-law dependence on frequency $f$ in a
large range: $P(f)\propto f^{-\alpha} $, where the index $\alpha$
depends on the intrinsic nature of the fluctuations. This is
related to the SF slope by $\alpha=1+\beta$.

\begin{figure}[t]
\begin{center}
\begin{tabular}{c}
{\resizebox{7cm}{!}{\includegraphics{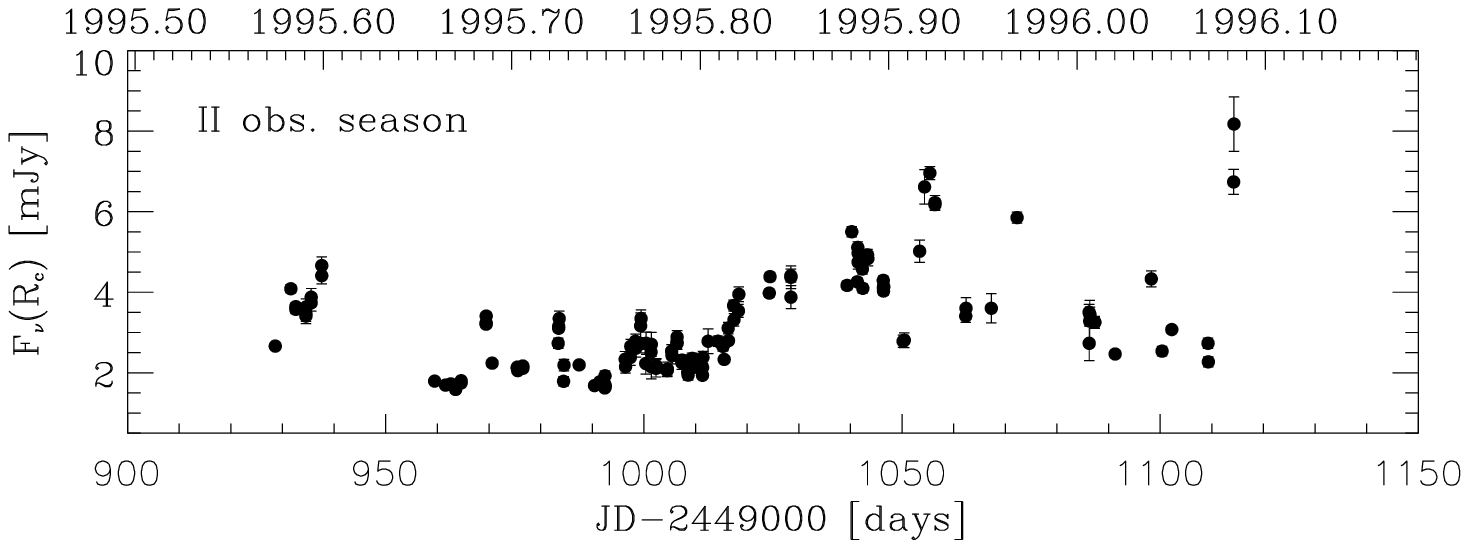}}} \\[-3mm]
{\resizebox{7cm}{!}{\includegraphics{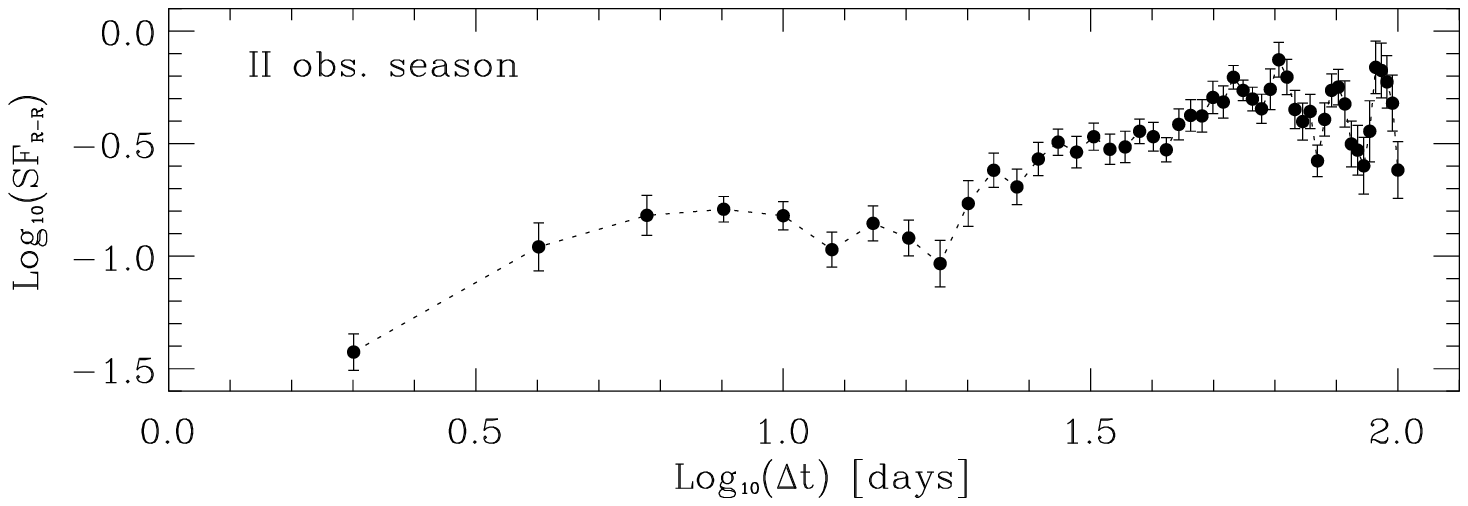}}} \\[-3mm]
{\resizebox{7cm}{!}{\includegraphics{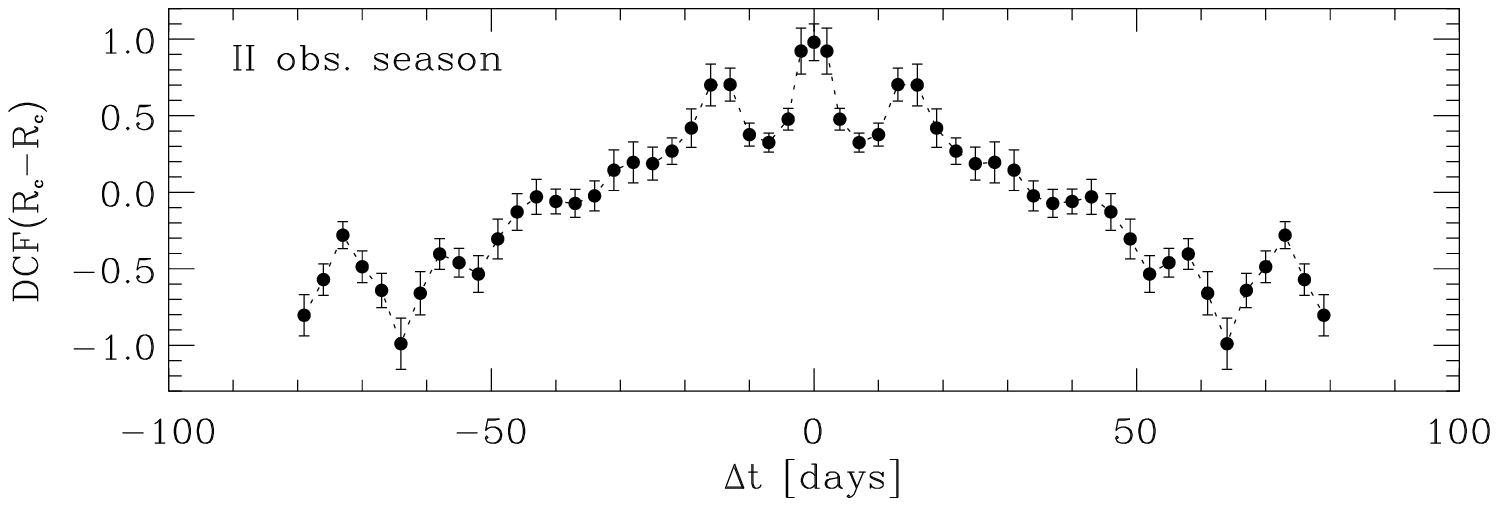}}} \\[-3mm]
{\resizebox{7cm}{!}{\includegraphics{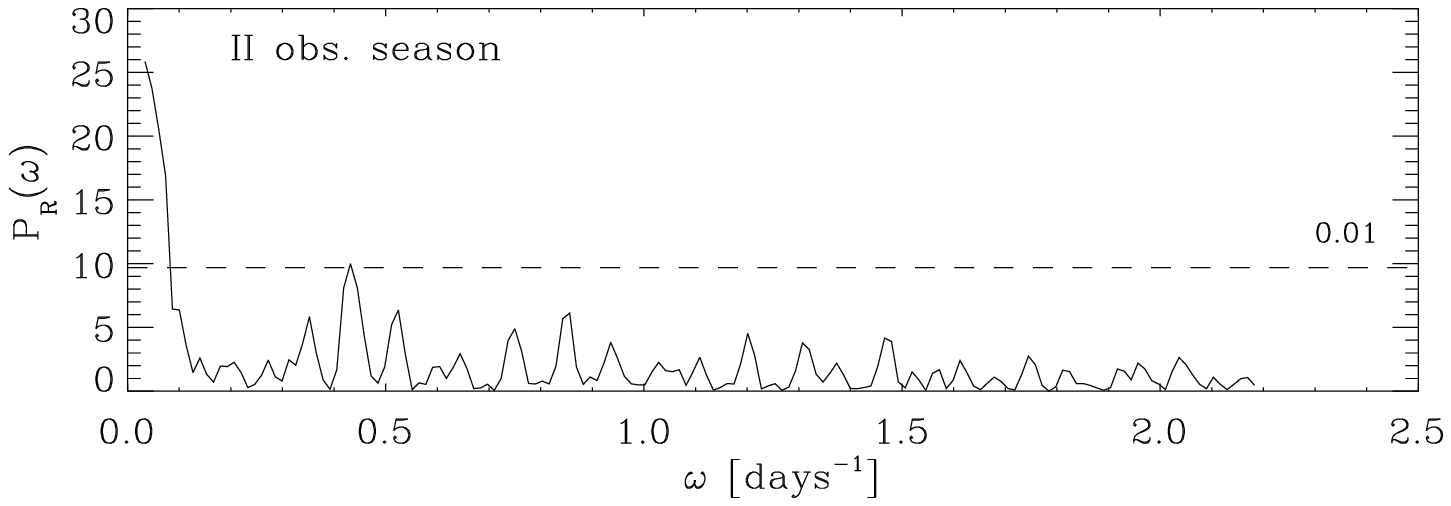}}} \vspace{-5mm}
\end{tabular}
\end{center}
\caption{The II observing season (July 1995 -- January 1996)
$R_{c}$ flux light curve, its structure function (data bin: 1 day,
SF bin: 2 days), discrete correlation function (data bin: 1 day,
DCF bin: 3 days), and periodogram ($\omega=2\pi f$). The dashed line
indicates the 99\% significance threshold.} \label{fig:sf-dcf95-96}
\end{figure}

The DCF may be used to auto- and cross-correlate unevenly sampled
data (see e.g.\ Edelson \& Krolik 1988; Hufnagel \& Bregman 1992).
This function does not require any data interpolation or
assumption about the light curve. The pairs $(a_{i},b_{j})$ of two
discrete datasets are first combined in unbinned discrete
correlations
\begin{equation}\label{eq:udcf}
  {\rm UDCF}_{ij}=\frac{(a_{i}-<a>)(b_{j}-<b>)}{\sigma_{a}\sigma_{b}},
\end{equation}
where $<a>$ and $<b>$ are the average values of the samples and
$\sigma_{a},\sigma_{b}$ their standard deviations. Each of these
correlations is associated with the pairwise lag $\Delta t_{ij} =
t_{j} - t_{i}$ and every value represents information about real
points. The correlations (\ref{eq:udcf}) do not give a proper
normalization if the data are noisy. The DCF is obtained by
binning the ${\rm UDCF}_{ij}$ in time for each time lag $\Delta
t$, and averaging over the number $M$ of pairs whose time lag
$\Delta t_{ij}$ is inside $\Delta t$, i.e.: ${\rm DCF}(\Delta
t)=1/M\sum_{ij} {\rm UDCF}_{ij}$. The choice of the bin size is a
free parameter, and is governed by trade-off between the desired
accuracy in the mean calculation, and the desired resolution in
the description of the correlation curve. A preliminary time
binning of data usually leads to better results, smoothing out
spurious spikes; also in this case the choice of the bin size is
determined by a balance between resolution and noise.

\begin{figure}[t]
\begin{center}
\begin{tabular}{c}
{\resizebox{7cm}{!}{\includegraphics{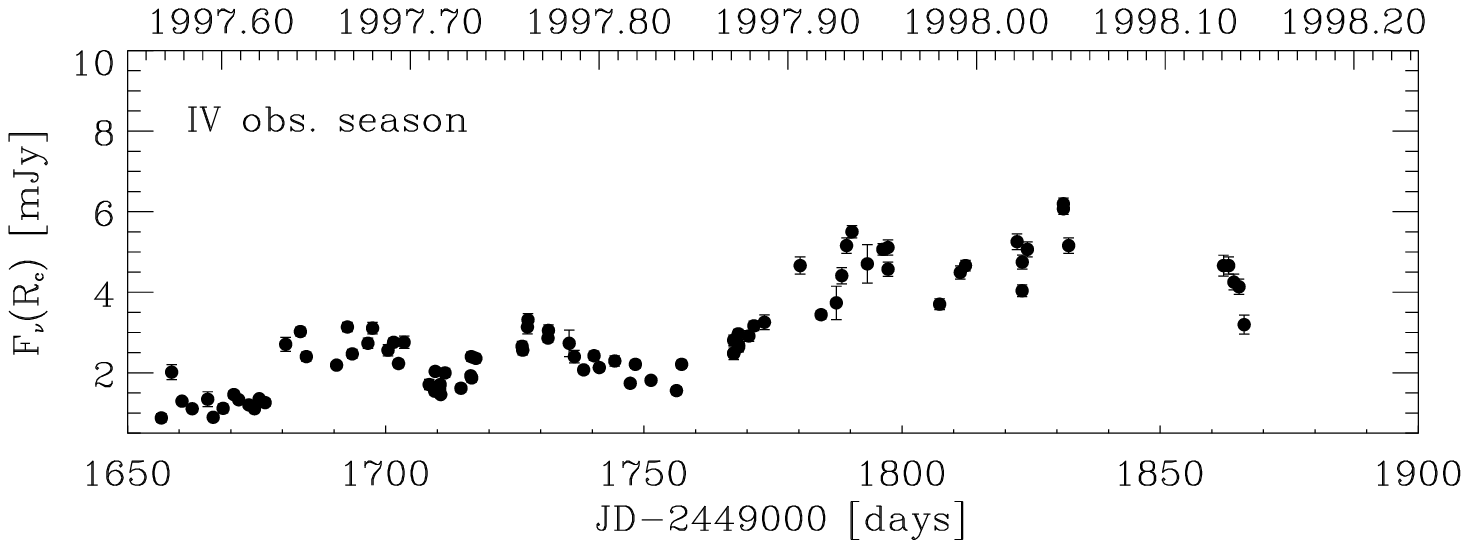}}} \\[-3mm]
{\resizebox{7cm}{!}{\includegraphics{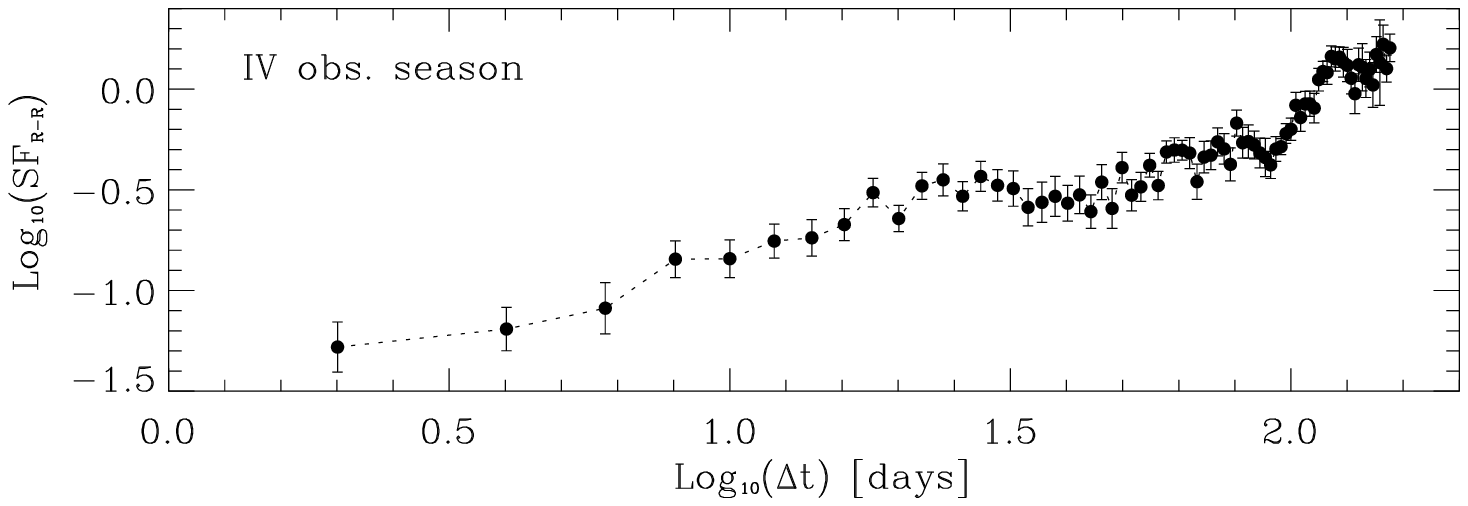}}} \\[-3mm]
{\resizebox{7cm}{!}{\includegraphics{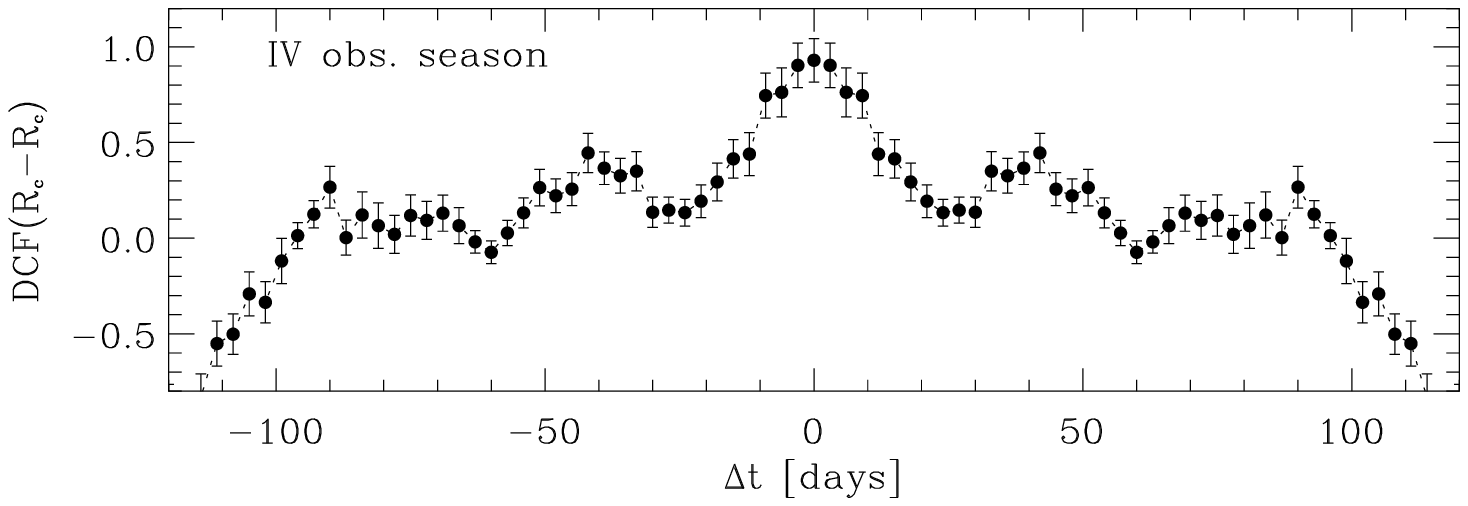}}} \\[-3mm]
{\resizebox{7cm}{!}{\includegraphics{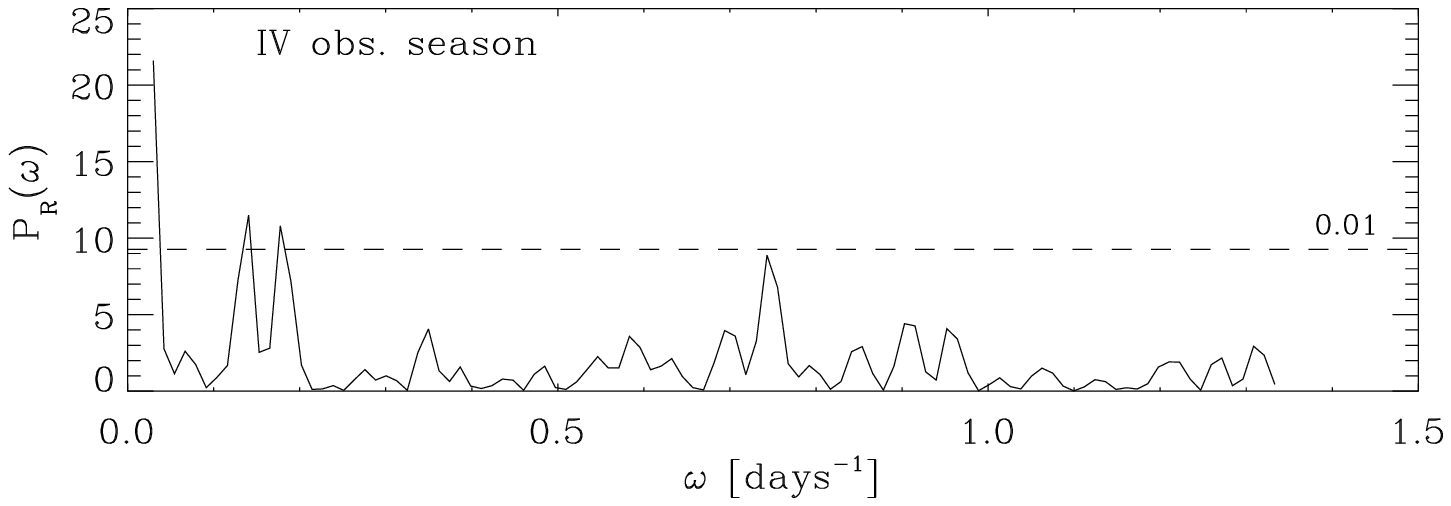}}} \vspace{-5mm}
\end{tabular}
\end{center}
\caption{The IV observing season (July 1997 -- February 1998)
$R_{c}$ flux light curve, its structure function (data bin: 1 day,
SF bin: 2 days), discrete correlation function (data bin: 1 day,
DCF bin: 3 days), and periodogram ($\omega=2\pi f$). The dashed line
indicates the 99\% significance threshold.}
\label{fig:sf-dcf97-98}
\end{figure}

The shapes of the SF and DCF reflect the nature of the process
underlying the data train. Uncorrelated data produce a ``white
noise" behaviour, characterized by a constant PSD ($\alpha=0$).
Shot noise or Brownian noise results from a sequence of random
pulses, i.e.\ a superimposition of similar shaped bumps randomly
distributed in time, with a long-term ``memory''. The PSD of the
standard shot noise has power law index around $\alpha=2$. Flicker
noise, called also ``pink noise'' or $1/f^\alpha$ noise (with
$\alpha$ approximatively in the range between 1 and 2), is an
halfway between the previous two ones. This could be the result of
parallel relaxation processes keeping up a certain level of memory
(Ciprini et al. 2003) .

The typical BL Lac variability seems to place in between the shot
and the flickering noise, even if different mechanisms could
explain the rapid intrinsic flickering and the long-term flux
oscillations.

\begin{figure}[t]
\begin{center}
\begin{tabular}{c}
{\resizebox{7cm}{!}{\includegraphics{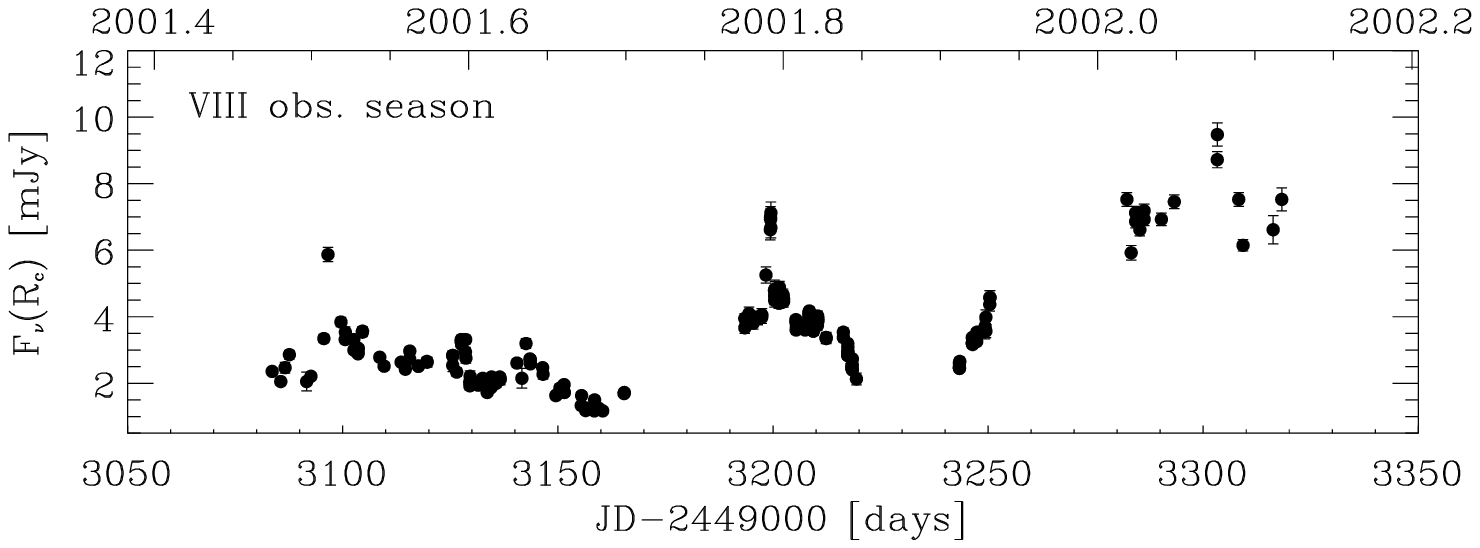}}} \\[-3mm]
{\resizebox{7cm}{!}{\includegraphics{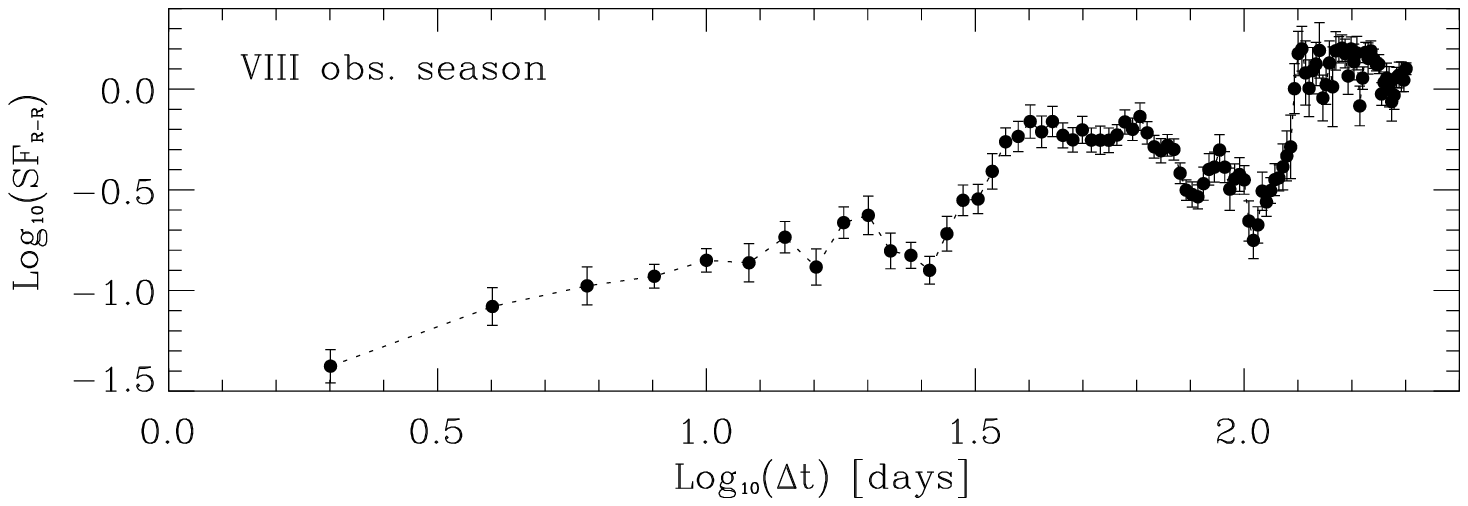}}} \\[-3mm]
{\resizebox{7cm}{!}{\includegraphics{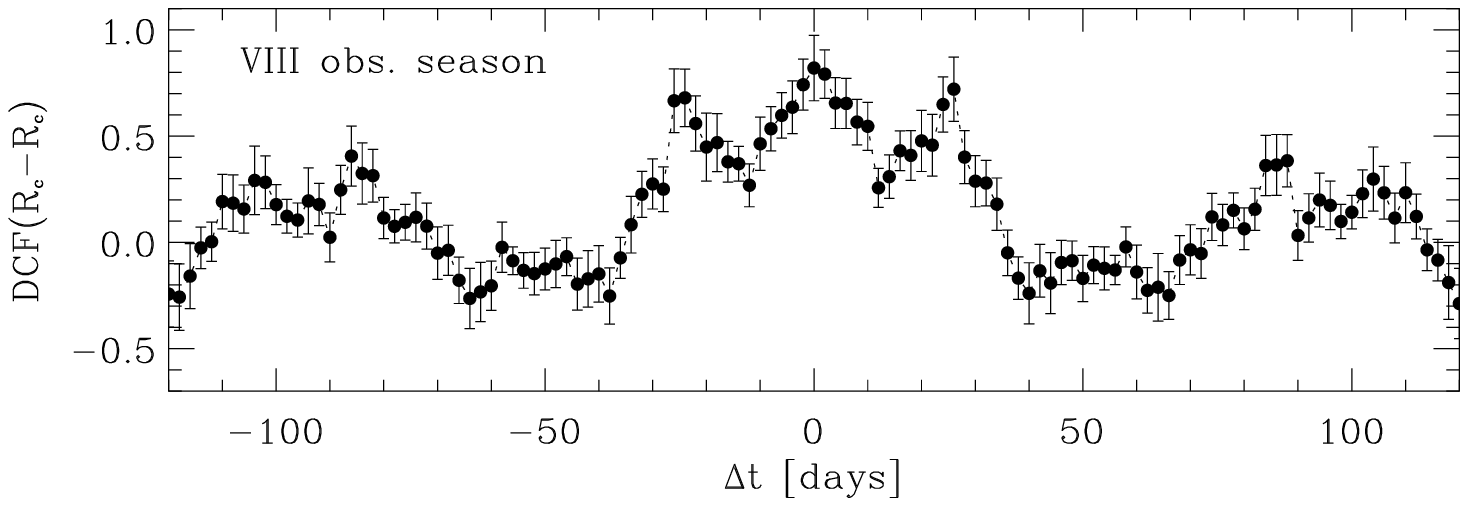}}} \\[-3mm]
{\resizebox{7cm}{!}{\includegraphics{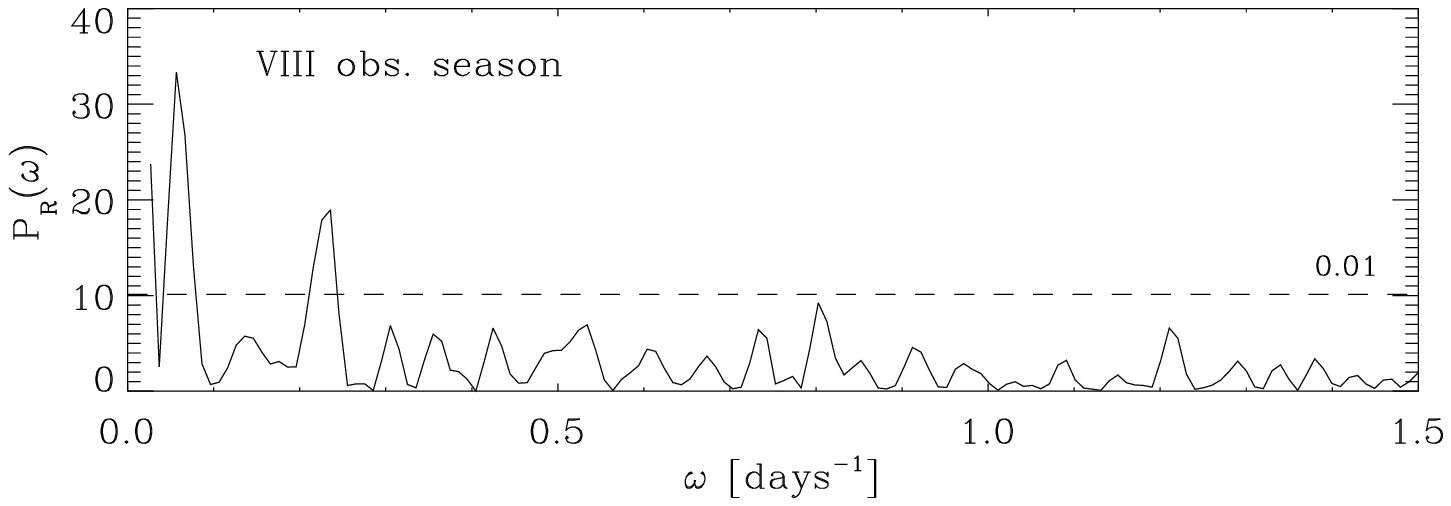}}} \vspace{-5mm}
\end{tabular}
\end{center}
\caption{The VIII observing season (June 2001 -- February 2002)
$R_{c}$ flux light curve, its structure function (data bin: 1 day,
SF bin: 2 days), discrete correlation function (data bin: 1 day,
DCF bin: 2 days), and periodogram ($\omega=2\pi f$). The dashed line
indicates the 99\% significance threshold.}
\label{fig:sf-dcf01-02}
\end{figure}
%
%
%
\begin{table*}[t!!]
\caption[]{Summary of all the time scales revealed by the
statistical analysis (when possible) performed with the structure
function (SF), the discrete correlation function (DCF), and the
Lomb-Scargle periodogram ($P(\omega)$). Columns report: (1)
observing period, (2) period length, (3) time scales estimated by
visual inspection, (4) time scales calculated by deep drops in the
SF, (5) SF slope, (6) time scales deduced from the SF transition
to the plateau, (7) time scales estimated from the DCF peaks, (8)
time scales derived from the peaks of the periodogram.}
\label{tab:timescalestable} \vspace{-0.2cm} \centering {}
\begin{tabular}{lccccccc}
\vspace{-3mm} \\
\hline \hline
\vspace{-3mm} \\
Obs.\ period &  Duration&  VE $T$ & SF $T_{dr}$ & SF slope &  SF $T_{to}$ & DCF $T_{pe}$ & P($\omega$) $T$ \\
      &  [days]  &  [days]         & [days]    & $\beta$ &       [days] & [days]       & [days] \\
\hline \hline
1976-2002$~(\dag)$   &  26y & ... & 320,1.2y,1.9y,6.5y    &  0.55$\pm$0.06 &  ...& 340,1.2y,2.1y & 1.0y,1.2y,2.0y,6.4y \\
1994-2002$~(\dag)$   &  7.3y         &  320, 1.2y  & 327,1.2y,1.9y  & 0.57$\pm$0.02  &   128 & 1.2y,1.9y &150,1.0y,1.9y \\
II season &  186     &  120           & 12,18,74       &0.68$\pm$0.06  &  ... & 13 & 14 \\
III season&  131     &  ...           &38,64       & 0.57$\pm$0.09  &  ...  & 68 & 70  \\
IV season &  210     &  45           & 130         & 0.70$\pm$0.04  &   122   & 42 & 35, 45 \\
V season &  230     & ...            & 54          & 0.72$\pm$0.08  &   ...    & 54 &... \\
VI season &  235     & ...            & 88          & ...  &   ...    &... &    87 \\
VII season&  220     &  40            & 40          &1.05$\pm$0.08  &   ...    & 38,95 & ...  \\
VIII season& 235     &  24           & 26, 104     & 0.73$\pm$0.09
&   40    & 26,88,104 & 27,111 \\ \hline \hline \end{tabular}
\begin{list}{}{} \item[$\dag $] Time scales followed by ``y'' are
expressed in years. \end{list}
\end{table*}

An analogous technique to the Fourier analysis for discrete
unevenly sampled data trains is the Lomb-Scargle periodogram
$P(\omega)$, useful to detect the strength of harmonic
components with a certain angular frequency $\omega=2\pi f$ (Lomb 1976; Scargle
1982; Horne \& Baliunas 1986).

We applied the SF, DCF and periodogram to the historical light
curve of GC 0109+224, and to each observing season of our best
sampled $R_{c}$ light curve. The observed magnitude was
transformed into flux density, corrected by the Galactic
absorption (derived by Schlegel et al.\ 1998, see table
\ref{tab:samplingprop}). Fluxes relative to zero-magnitude values
were taken from Bessel (1979).

A summary of the clearest statistical outcomes of our calculations
is reported in Table \ref{tab:timescalestable}: we extracted the
most reliable characteristic time scales from the applied methods
as well as from a preliminary visual inspection. The visual time
scales are derived by identifying the repetitions of the flare
peaks with similar shape, or the interval among the minima, in
case we have recognized an alternating shape, or a train of
successive pulses. In the time intervals where the SF slope is
distinguishable in the log-log plots, we calculated its power
index $\beta$ trough a linear regression.

No reliable feature was obtained from the analysis of the complete
historical 1906--2002 $B$ light curve (an observational period of
about 9400 days), due to the poor and very irregular sampling. In
the 1976--2002 better sampled light curve, characteristic time
scales of about 320 days and 1.2, 1.9, and 6.5 years are derived
from SF minima, DCF peaks, and periodogram peaks. The first three
time scales are confirmed by the analysis of the 1994--2002
$R_{c}$ light curve, where a significant slope of $0.57\pm0.02$
was calculated from the SF, which shows more clearly a plateau
beginning from $\Delta t=128$ days (Fig. \ref{fig:sf-dcf94-02}).
The 1.0 and 2.0 year time scales might be spurious scales due to
the seasonal interruption in the sampling.

In the following we discuss briefly the single observing seasons,
investigating characteristic time scales from one day to some
months. Spurious correlations due to external factors are more
easily avoided and the conclusions are more significant (where the
sampling is sufficient) because of the lack of long empty windows
among the observations and of a more regular sampling.

During the July 1995 -- January 1996 period (II season, Figure
\ref{fig:sf-dcf95-96}) we revealed a SF slope of
$\beta=0.68\pm0.06$, and the DCF profile showed a clear shot noise
feature, corresponding to the couple of months oscillation of the
base level flux. A variability time scale of 12--14 days comes out
from SF, DCF, and periodogram. In the October 1996 -- February
1997 (III season), the SF did not exhibit a clear sign of a
plateau. The DCF showed a central shot noise broad peak and two
prominent peaks at $68$ days, confirmed by the drop in the SF and
by the peak of the periodogram. In the period between July 1997 --
February 1998 (IV season, Fig. \ref{fig:sf-dcf97-98}), the SF
showed a steep slope ($\beta=0.70\pm 0.04$), with hints of an
upper flattening again at about 120 days. This agrees with the
shot noise main component of the DCF, spreading for 120 days, and
reflects the nearly monotonic rise of the underlying flux in the
considered period. The periodogram shows two modest peaks
corresponding to time scales of 35 and 45 days; this second value
is weakly confirmed by the DCF, while the SF presents a minimum at
twice this time.

During the July 1998 -- February 1999 observing period (V season),
a time scale of about 54 days is found. The SF slope is similar to
that of the previous season. In the June 1999 -- February 2000
period (VI season), a typical time scale of about 8 days comes out
from a SF drop and periodogram analysis. The July 2000 -- February
2001 observational window (VII season), the SF displayed a steep
slope ($\beta=1.05\pm 0.09$), with no identifiable flattening.
Visual inspection, SF drop, and the highest DCF peak suggest a
scale of about 40 days.

In the last observing window (VIII season, June 2001 -- February
2002, Figure \ref{fig:sf-dcf01-02}) a good sampling was obtained
and the characteristic time scales can be clearly extrapolated.
The SF shows a slope $\beta=0.73\pm 0.10$, with a flattening at 40
days. The DCF is characterized by a central shot noise broad peak
of about 40 days, with a noticeable peak at about 26
days, and minor peaks at 88 and 104 days. Such time scales are
confirmed by the minima of the SF and the peaks of the
periodogram.

From the above discussion and from the results summarized in Table
\ref{tab:timescalestable}, one can conclude that the light curves
of the BL Lac object GC 0109+224 show indications of recurrent
time scales of variability, from a dozen days to a few years.
Scales of about 25--40 days and 1.2, 1.9 years were found several
times, and the time scale of 6.4 years is similar to the variability
period recognized in another BL Lac object: AO 0235+16 (Raiteri et
al.\ 2001). Slopes in the SF were reliably determined, but the SF
plateau can be recognized only for the best sampled light curves.
The values of $\beta$ imply a variability mode between the
flickering and the shot noise.

\section{Summary}
Seven years of optical monitoring of the BL Lac object GC 0109+224
have confirmed its intense variability. The source showed
variations of about 2.5 magnitudes in less than one year, and a
moderate-amplitude continuous flaring. This flaring activity
appears superposed to a long-term oscillation of the base-level
flux, which was recognized to vary with a time scale of about 11.6
years (Smith \& Nair 1995).

The mode of variability is intermittent, characterized by
relatively fast drops of the flux and by a not regular alternation
of semi-quiescent and bursting phases. Our findings suggest that
also in the GC 0109+224 as in other objects (like BL Lac, Villata
et al.\ 2002) we are seeing the results of two (or maybe more)
distinct mechanisms playing on different time scales. Again
long-term variations seem to be essentially achromatic, while
short-term flares at least in some cases follow the ``flatter when
brighter'' feature. Indeed, when plotting the optical spectral
index versus flux during three well-defined flares, clear
hysteresis loops result (as usually found when analyzing the flux
behaviour of BL Lacs at X-ray bands), possibly indicating that
radiative cooling of a single population of accelerated electrons
in a jet flaring region is dominating.

Several typical variability timescales appear from the
quantitative analysis of our data, from a dozed days to a few
years, but our sampling is not good enough to display any
convincing evidence of periodicity. The longest characteristic
time scale seen in the flux changes of GC 0109+224 is 6.5 years,
very similar to the period recognized in AO 0235+16 (Raiteri et
al. 2001), but poorly evident in this case. A further, continuous
monitoring is required to see whether some of the time scales
found in the present work are effectively linked to variability
recurrent mechanisms. The optical variability of GC 0109+224 seems
characterized by the $1/f^{\alpha}$ behaviour (with $1.57 < \alpha
< 2.05$), meaning a fluctuation mode in between the flickering and
the shot noise, a common feature in blazars. The power law decline
of the PSD means that the frequency occurrence of a specific
variation is inversely proportional to the strength of the
variation. This could be caused by stochastic relaxation processes
(Ciprini et al. 2003), in operation on the baseline or
parsec-scale jet and generating the intermittent optical
behaviour. Such intrinsic flux flickering appear then superposed
to long term trends, maybe produced by external factors like jet
bending.

We finally remark that the optical behaviour, the high
polarization, the X-ray loudness (that could suggest a possible
positive detection of $\gamma$-ray emission for the next
generation of satellites like Agile and Glast), and the probable
no low redshift, make GC 0109+224 an interesting BL Lac object. In
this picture our data represent a useful database of the recent
optical history of GC 0109+224, for future optical and
multifrequency researches.

\begin{acknowledgements}
We greatly appreciate the anonymous referee, for his helpful
comments and suggestions. We thank also Dr. M. Fiorucci for some
useful discussions, and the past collaborators who obtained and
reduced part of the data. This research has made use also of:
\begin{itemize}
  \item the NASA/IPAC Extragalactic Database (NED), which is operated by
the JPL-Caltech, under contract with NASA;
\item Simbad Astronomical Database and Vizie-R Catalogues Services, operated at Centre de
Données astronomiques de Strasbourg (CDS), France.
\end{itemize}
This work was partly supported by the Italian Ministry for University and
Research (MURST) under grant Cofin 2001/028773.

\end{acknowledgements}

\bibliographystyle{aa}

\end{document}